\documentclass[journal,a4paper,10pt]{IEEEtran}

\usepackage[T1]{fontenc}
\usepackage[latin9]{inputenc}

\usepackage{epsfig,psfrag}
\usepackage{amsmath,amsfonts,amssymb,amsxtra,bm}
\usepackage{graphicx}
\usepackage{color}
\usepackage{cite}
\usepackage{subfigure}

\newcommand{\eq}{\,=\,}
\newcommand{\be}{\begin{equation}}
\newcommand{\ee}{\end{equation}}
\newcommand{\ist}{\hspace*{.3mm}}
\newcommand{\rmv}{\hspace*{-.3mm}}

\allowdisplaybreaks

\begin{document}

\title{Likelihood Consensus and Its Application to Distributed Particle Filtering\thanks{Copyright (\copyright) 2012 IEEE. Personal use of this material is permitted. 
However, permission to use this material for any other purposes must be obtained from the IEEE by sending a request to pubs-permissions@ieee.org.}
\thanks{Ondrej Hlinka, Ondrej Slu\v ciak, Franz Hlawatsch, and Markus Rupp are with the Institute of 
Telecommunications, Vienna University of Technology, A-1040 Vienna, Austria (e-mail: \{ondrej.hlinka,$\ist$ondrej.sluciak, franz.hlawatsch,$\ist$markus.rupp\}@nt.tuwien.ac.at). 
Petar M.\ Djuri\'c is with the Department of Electrical and Computer Engineering, Stony Brook University, 
Stony Brook, NY, USA (e-mail: djuric@ece.sunysb.edu). This work was supported by the FWF under Awards S10603 and S10611 within the National Research 
Network SISE, by the NSF under Award CCF-1018323, and by the ONR under Award N00014-09-1-1154. Parts of this work were previously presented
at the 44th Asilomar Conf. Sig., Syst., Comp., Pacific Grove, CA, Nov.\ 2010 and at IEEE ICASSP 2011, Prague, Czech Republic, May 2011.
The simulation source files are available online at http://ieeexplore.ieee.org.
}\vspace*{3mm}}

\author{{\it Ondrej Hlinka}, Student Member, IEEE, 
{\it Ondrej Slu\v ciak}, Student Member, IEEE, 
{\it Franz Hlawatsch}, Fellow, IEEE, 
{\it Petar M.\ Djuri\'c}, Fellow, IEEE,
and {\it Markus Rupp}, Senior Member, IEEE\\[3mm]
\vspace*{2mm}
 }


\maketitle

\vspace*{-5mm}

\begin{abstract}
We consider distributed state estimation in a wireless sensor network without a fusion center. 
Each sensor performs a global estimation task---based on the past and current measurements of \emph{all} sensors---using 
only local processing and local communications with its neighbors. In this estimation 
task, the joint (all-sensors) likelihood function (JLF) plays a central role as it epitomizes the measurements of all sensors.
We propose a distributed method for computing, at each sensor,  
an approximation of the JLF by means of consensus algorithms. 
This ``likelihood consensus''  method is applicable if the local likelihood functions of the various sensors (viewed as 
conditional probability density functions of the local measurements)
belong to the exponential family of distributions. 
We then use the likelihood consensus method to implement a distributed particle filter and a distributed Gaussian particle filter. 
Each sensor runs a local particle filter, or a local Gaussian particle filter, that computes a global state estimate. 
The weight update in each local (Gaussian) particle filter employs  
the JLF, which is obtained through the likelihood consensus scheme. 
For the distributed Gaussian particle filter, 
the number of particles can be significantly reduced by means of an additional consensus scheme. 
Simulation results are presented to assess  the performance of the proposed 
distributed particle filters for a multiple target tracking problem. 
\end{abstract}


\section{Introduction}

Distributed estimation in wireless sensor networks has received significant attention recently 
(e.g., \cite{zhao2004wsn, haykin2009handbook, ferrari2010sensor}). Applications include
machine and structural health monitoring, pollution source localization, habitat monitoring, and target tracking.
Typically, a wireless sensor network is composed of 
battery-powered sensing/processing nodes---briefly called ``sensors'' hereafter---which possess limited sensing, computation, and communication capabilities.
 
Centralized estimation techniques 
transmit sensor data 
to a possibly distant fusion center 
\cite{zhao2004wsn}. This may require energy-intensive communications over large distances 
or complex multi-hop routing protocols, 
which results in poor scalability. 
Centralized techniques are also less robust, 
and less suitable if the estimation results have to be available at the sensors (e.g., in sensor-actuator networks \cite{nayak2010wireless}). 
Furthermore, the fusion center must be aware of the measurement models and, possibly, additional parameters of all sensors. 
By contrast, decentralized estimation techniques without a fusion center 
use in-network processing and neighbor-to-neighbor communications to achieve low energy consumption as well as high robustness 
and scalability. The sensors do not require knowledge of the network topology, and no routing protocols are needed.

There are two basic categories of decentralized estimation techniques. In the first, information is transmitted in a sequential manner 
from sensor to sensor 
\cite{ zhao2007idd, zhao2007dsb, hlinka2009time}. In the second, 
each sensor diffuses its local information 
in an iterative process using broadcasts to a set of neighboring sensors (e.g., \cite{baraniuk2006esa}). This second category is more robust but involves an increased communication overhead. It includes 
consensus-based estimation techniques, which 
use 
distributed algorithms for reaching a consensus (on a sum, average, maximum, 
etc.)\ in the network \cite{olfati2005distributed,dimakis2010gossip}. 
Examples 
are gossip 
algorithms \cite{dimakis2010gossip}, 
consensus algorithms \cite{OlfatiSaber07consensus}, 
and combined approaches \cite{Aysal09broadcast}. 

In this paper, we consider a decentralized wireless sensor network architecture without a fusion center and
use consensus algorithms to perform a \emph{global} estimation task through \emph{local} processing and communications, 
in a way such that the final global estimate is available locally at each sensor.
(``Global'' estimation 
means that the measurements of \emph{all} sensors 
are processed by each sensor.) This can be based on the joint (all-sensors) likelihood function, abbreviated JLF, which epitomizes the measurements of all sensors. 
The JLF is then required to be known by all sensors. For example, a global particle filter (PF) \cite{gordon1993novel,doucet2001sequential,arulampalam2002tpf} 
that processes all sensor measurements relies on the pointwise evaluation of the JLF to perform its weight update.

The main contribution of this paper is a 
distributed method for calculating the JLF or an approximation of the JLF at each sensor. 
Generalizing our previous work 
in \cite{hlinka2010likelihoodcons, hlinka2011dgpf}, this method is suited to sensors with local likelihood functions 
that are members of the exponential family of distributions.
A consensus algorithm---calculating sums---is used for a decentralized, 
iterative computation of a sufficient statistic that describes the (approximate) JLF as a function of the state to be estimated. 
Consequently, we refer to our method as \emph{likelihood consensus} (LC). 
The LC scheme requires communications only between neighboring sensors 
and operates without routing protocols. 
We furthermore propose an application of our LC method in a distributed PF scheme and in a distributed Gaussian PF scheme. 
Each sensor runs a local PF (or a local Gaussian PF \cite{kotecha2003gaussian}) that computes a global state estimate incorporating all sensor measurements. 
At any given PF recursion, each local (Gaussian) PF draws a set of particles and updates their weights based on an evaluation of the JLF at these particles. 
For the distributed Gaussian PF, 
the number of particles employed by each local Gaussian PF 
can be significantly reduced 
by means of a second consensus scheme. 

Alternative consensus-based distributed PF schemes 
have been 
proposed in  
\cite{farahmand2011set,gu2007distributed,gu2008consensus,oreshkin2010async,mohammadi2011consensus,xaver2011localization}. The method 
described\linebreak 
in  \cite{farahmand2011set} uses one consensus algorithm per particle to calculate products of local particle weights. 
To reduce the 
communication requirements, 
the number of particles is kept small by an 
adaptation of the proposal distribution.
Nevertheless, the number of consensus algorithms required 
can be significantly higher than in our approach. Furthermore, the random number generators of the individual sensors must be synchronized. 
On the other hand, since no approximation of the JLF is required, 
the performance can be closer to that of a centralized PF. 
The consensus-based distributed PFs proposed in \cite{gu2007distributed} and \cite{gu2008consensus} 
rely on local PFs that update their weights using only the 
\emph{local} likelihood functions instead of the JLF. 
Gaussian or Gaussian mixture approximations of local posteriors are then computed, 
and a consensus algorithm is 
used to fuse 
these approximations. 
However, this fusion rule is suboptimal and leads to a performance loss. 
In \cite{oreshkin2010async}, a novel gossiping approach implementing 
an approximation of the optimal fusion rule is employed to 
construct a Gaussian approximation of the global posterior. 
However, again only local likelihood functions are used by the local PFs, 
and the estimation performance is worse than in our approach. 
In \cite{mohammadi2011consensus}, a distributed unscented PF is proposed that uses local measurements for proposal adaptation and an optimal consensus-based fusion rule to compute global estimates from local estimates. The distributed PF proposed in \cite{xaver2011localization} operates across clusters of sensors and uses a modified maximum consensus algorithm to aggregate the local posterior distributions from all clusters. 

Distributed PFs that do not rely on consensus algorithms have been presented in \cite{coates2004distributed,sheng2005distributed,hlinka2009timePF}.
In these methods, a path through the sensor network is adaptively determined 
by means of a decentralized sensor scheduling 
algorithm. 
Parametric representations of partial likelihood functions or of partial posteriors are transmitted along this path. The last sensor in the path 
obtains the complete global information and is thus able to compute a global estimate. 
In general, these methods are not as robust to sensor failure as the consensus-based methods.  
However, in certain applications, their communication requirements may be much lower. 

This paper is organized as follows. 
In Section \ref{sec:system-model_exp}, we describe 
the system model and 
review 
sequential Bayesian estimation. 
To prepare the ground for the 
LC method, an approximation of the exponential class of distributions is discussed in Section \ref{sec:approx-exp}. 
The LC method is presented in Section \ref{sec:likelihood-cons_exp}. 
In Section \ref{sec:gauss}, we consider the special case of additive Gaussian measurement noise.
The application of LC to distributed particle filtering and distributed Gaussian particle filtering is considered in Section \ref{sec:distributed-particle-filtering} 
and \ref{sec:distributed-gaussian-particle-filtering}, respectively.
Finally, in Section \ref{sec:sim-results}, the proposed distributed PFs are applied to 
multiple target tracking, and simulation results are presented.


\vspace{-.5mm}

\section{System Model and Sequential Bayesian Estimation}\label{sec:system-model_exp}

\vspace{.5mm}

We consider a wireless sensor network consisting of $K$ sensors. At a given discrete time $n$, each sensor estimates
a global $M$-dimensional state 
$\mathbf{x}_{n}=(x_{n,1}\cdots\ist\ist x_{n,M})^\top\! \in \mathbb{R}^M\rmv$ based on 
all sensor 
measurements. 
The state evolves 
according to the state-transition probability density function (pdf) 
$f(\mathbf{x}_{n}|\mathbf{x}_{n-1})$. At time $n$, the $k\ist$th sensor ($k \!\in\! \{1,\dots,K\}$) acquires an $N_{n,k}$-dimensional measurement 
$\mathbf{z}_{n,k} \in \mathbb{R}^{N_{n,k}}\rmv$. 
The 
relationship between $\mathbf{z}_{n,k}$ and 
$\mathbf{x}_n$ is described by the \emph{local likelihood 
function}\footnote{The 
notation $f(\mathbf{z}_{n,k}|\mathbf{x}_n)$ suggests that $\mathbf{x}_n$ is a random vector. However, for the LC method to be 
presented in Section \ref{sec:likelihood-cons_exp}, $\mathbf{x}_n$
is also allowed to be deterministic, in which case the notation $f(\mathbf{z}_{n,k};\mathbf{x}_n)$ would be more 
appropriate.} 
$f(\mathbf{z}_{n,k}|\mathbf{x}_n)$, and the 
relationship between 
the all-sensors measurement vector $\mathbf{z}_{n} \!\triangleq\! (\mathbf{z}_{n,1}^{\top} \!\cdots\ist \mathbf{z}_{n,K}^{\top})^{\top}\!$ and 
$\mathbf{x}_n$ is described by the JLF $f(\mathbf{z}_n|\mathbf{x}_n)$. All $\mathbf{z}_{n,k}$ are assumed conditionally independent given $\mathbf{x}_n$, so that
the JLF is the product of all local likelihood functions, i.e., 
\be
f(\mathbf{z}_{n}|\mathbf{x}_n) \eq \prod_{k=1}^{K}f(\mathbf{z}_{n,k}|\mathbf{x}_n) \,.
\vspace{1mm}
\label{eq:joint_likelihood_prod}
\ee
We write $\mathbf{z}_{1:n} \!\triangleq\! ( \mathbf{z}_{1}^{\top} \!\cdots\ist \mathbf{z}_{n}^{\top} )^\top\rmv$ for the vector of the 
measurements of all 
sensors up to time $n$.

In the sequel, we will use the following 
assumptions. 
First, the current state $\mathbf{x}_{n}$
is conditionally independent of all past measurements, $\mathbf{z}_{1:n-1}$, given the previous state $\mathbf{x}_{n-1}$, i.e.,
\begin{equation}
f(\mathbf{x}_{n}|\mathbf{x}_{n-1},\mathbf{z}_{1:n-1}) \,=\, f(\mathbf{x}_{n}|\mathbf{x}_{n-1}) \,.
\label{eq:stateTransitionPdf}
\end{equation}
Second, the current measurement $\mathbf{z}_{n}$ is conditionally independent of all past measurements, $\mathbf{z}_{1:n-1}$,
given the current state $\mathbf{x}_{n}$, 
\vspace{-1.5mm}
i.e.,
\begin{equation}
f(\mathbf{z}_{n}|\mathbf{x}_{n},\mathbf{z}_{1:n-1}) \,=\, f(\mathbf{z}_{n}|\mathbf{x}_{n}) \,.
\label{eq:likelihood_network}
\end{equation}
Finally, sensor $k$ knows the state-transition pdf $f(\mathbf{x}_{n}|\mathbf{x}_{n-1})$ and its own local likelihood function $f(\mathbf{z}_{n,k}|\mathbf{x}_n)$ 
as well as the pdf $f(\mathbf{x}_0)$ of the initial state $\mathbf{x}_0$, but 
it does not know the local likelihood functions of the other sensors, i.e., $f(\mathbf{z}_{n,k'}|\mathbf{x}_n)$ for $k'\!\not=\! k$.

We briefly review sequential Bayesian state estimation \cite{ristic2004bkf}, which will be considered 
as a motivating application of the LC method.
At time $n$, 
each sensor 
estimates the  
current state $\mathbf{x}_n$ from the 
measurements of all sensors up to time $n$, $\mathbf{z}_{1:n}$. 
For this task, we will use the minimum mean-square error (MMSE) estimator \cite{kay1998fundamentals},
\be
\label{eq:mmse_est}
\hat{\mathbf{x}}_{n}^{\text{MMSE}} \,\triangleq\,\ist \text{E}\{\mathbf{x}_{n}|\mathbf{z}_{1:n}\}
\,= \int \rmv\mathbf{x}_{n} \ist f(\mathbf{x}_{n}|\mathbf{z}_{1:n}) \, d\mathbf{x}_{n} \,,
\ee
which is implemented at each sensor. 
Here, a major problem---even in a 
centralized scenario---is to calculate the posterior pdf $\rmv f(\mathbf{x}_{n}|\mathbf{z}_{1:n})$. 
Using \eqref{eq:stateTransitionPdf} and \eqref{eq:likelihood_network}, 
the current posterior $f(\mathbf{x}_{n}|\mathbf{z}_{1:n})$ 
can be obtained sequentially from the previous posterior $f(\mathbf{x}_{n-1}|\mathbf{z}_{1:n-1})$
and the JLF $f(\mathbf{z}_{n}|\mathbf{x}_{n})$ by means of the following temporal 
recursion \cite{ristic2004bkf}:
\be
\label{eq:sequ_post_update}
f(\mathbf{x}_n|\mathbf{z}_{1:n}) \eq \frac{ f(\mathbf{z}_{n}|\mathbf{x}_{n})\int \rmv f(\mathbf{x}_{n}|\mathbf{x}_{n-1}) \ist f(\mathbf{x}_{n-1}|\mathbf{z}_{1:n-1}) \ist d\mathbf{x}_{n-1} }{ f(\mathbf{z}_n|\mathbf{z}_{1:n-1})} \,.
\ee 
However, for nonlinear/non-Gaussian cases, the computational complexity of 
sequential MMSE state estimation as given by \eqref{eq:mmse_est} and \eqref{eq:sequ_post_update} is typically prohibitive.
A computationally feasible approximation is provided by the PF \cite{doucet2001sequential,arulampalam2002tpf,ristic2004bkf}.
In a PF, the (non-Gaussian) posterior $f(\mathbf{x}_n|\mathbf{z}_{1:n})$ is represented by a set of samples (or particles) 
$\mathbf{x}_n^{(j)}\rmv$, $j=1,\dots,J\rmv$ and corresponding weights $w_{n}^{(j)}\rmv$.

As can be seen from \eqref{eq:mmse_est} and \eqref{eq:sequ_post_update}, obtaining the global estimate $\hat{\mathbf{x}}_{n}^{\text{MMSE}}$ at each sensor presupposes that each sensor knows the JLF $f(\mathbf{z}_{n}|\mathbf{x}_{n})$ as a function of the state $\mathbf{x}_{n}$ ($\mathbf{z}_{n}$ is observed and thus
fixed, and 
$f(\mathbf{x}_{n-1}|\mathbf{z}_{1:n-1})$ used in \eqref{eq:sequ_post_update} 
was calculated by each sensor at the previous time 
$n \!-\! 1$). 
In particular, a PF approximation of $\hat{\mathbf{x}}_{n}^{\text{MMSE}}$ 
relies on the pointwise evaluation of the JLF at the 
particles $\mathbf{x}_n^{(j)}$---i.e., on the evaluation of  $f(\mathbf{z}_{n}|\mathbf{x}_{n}^{(j)})$---to 
obtain the weights $w_{n}^{(j)}\rmv$. Since each sensor knows only its local likelihood function $f(\mathbf{z}_{n,k}|\mathbf{x}_{n})$, we need a
distributed method for calculating the JLF at each sensor. Such a method 
is proposed in Section \ref{sec:likelihood-cons_exp}. 

It is important to note that, although we consider distributed sequential Bayesian estimation and distributed particle filtering
as a motivating application, the proposed method can also be used for other distributed statistical inference tasks that require 
the pointwise evaluation of the JLF at the individual sensors.

\section{Approximation of the Joint Likelihood Function}\label{sec:approx-exp}

\vspace{.5mm}

The LC method can always be used if the local likelihood functions (viewed as 
conditional pdfs of the local measurements)
belong to the exponential family of distributions.
Typically, it requires an approximation of the local likelihood functions, and consequently of the JLF,
which 
is discussed in the following. 
In Section \ref{sec:LC-exact}, we will consider a class of JLFs for which an approximation is not needed.

\subsection{Exponential Family}

\vspace{.8mm}

In this paper, except in Section \ref{sec:LC-exact}, we assume that the local likelihood function of each sensor (viewed as the conditional pdf of $\mathbf{z}_{n,k}$) 
belongs to the exponential family of distributions \cite{bishop2006pattern}, 
i.e., 
\begin{align}
&f(\mathbf{z}_{n,k}|\mathbf{x}_n) \eq c_{n,k}(\mathbf{z}_{n,k}) \ist\exp\rmv\rmv \big(\mathbf{a}_{n,k}^{\top}(\mathbf{x}_n) \ist \mathbf{b}_{n,k}(\mathbf{z}_{n,k}) \nonumber\\[1mm]
&\rule{30mm}{0mm}- d_{n,k}(\mathbf{x}_n)\big) \,,  \quad\; k = 1,\ldots,K \ist,
\label{eq:loc_likelihood_exp}
\end{align}
with some time- and sensor-dependent functions $c_{n,k}(\cdot) \!\in\! \mathbb{R}_+$, $\mathbf{a}_{n,k}(\cdot) \!\in\! \mathbb{R}^q$, $\mathbf{b}_{n,k}(\cdot) \!\in\! \mathbb{R}^q$, 
and $d_{n,k}(\cdot) \!\in\! \mathbb{R}_+$, with arbitrary $q \!\in\! \mathbb{N}$.
We furthermore assume that sensor $k$ knows its own functions $c_{n,k}(\cdot)$, 
$\mathbf{a}_{n,k}(\cdot)$, $\mathbf{b}_{n,k}(\cdot)$, and $d_{n,k}(\cdot)$, 
but not $c_{n,k'}(\cdot)$, $\mathbf{a}_{n,k'}(\cdot)$, $\mathbf{b}_{n,k'}(\cdot)$, and $d_{n,k'}(\cdot)$ 
for $k'\!\not=\! k$. 
Using \eqref{eq:joint_likelihood_prod}, 
the JLF is obtained as
\begin{align}
f(\mathbf{z}_n|\mathbf{x}_n) &\eq \prod_{k=1}^{K} c_{n,k}(\mathbf{z}_{n,k}) \ist\exp\rmv \big(\mathbf{a}_{n,k}^{\top}(\mathbf{x}_n) \ist 
  \mathbf{b}_{n,k}(\mathbf{z}_{n,k}) \nonumber\\[-2mm]
&\rule{45mm}{0mm}- d_{n,k}(\mathbf{x}_n)\big) \label{eq:joint_likelihood_exp_0}\\[1mm]
& \eq C_n(\mathbf{z}_n) \exp\rmv\rmv\big( S_n(\mathbf{z}_n,\mathbf{x}_n) \big) \ist , 
\label{eq:joint_likelihood_exp}
\end{align}
where 
\be
C_n(\mathbf{z}_n) \,\triangleq\, \prod_{k=1}^{K} c_{n,k}(\mathbf{z}_{n,k})
\label{eq:joint_likelihood_exp_C}
\vspace{-2.5mm}
\ee
and
\vspace{-.5mm}
\be
S_n(\mathbf{z}_n,\mathbf{x}_n) \,\triangleq\,\rmv \sum_{k=1}^{K}\big[ \mathbf{a}_{n,k}^{\top}(\mathbf{x}_n) \ist \mathbf{b}_{n,k}(\mathbf{z}_{n,k}) - d_{n,k}(\mathbf{x}_n) \big] \,.
\label{eq:sum_Sz_exp}
\vspace{1mm}
\ee
Note that the JLF 
(viewed as the conditional pdf of $\mathbf{z}_{n}$) 
also belongs to the exponential family. 
The normalization factor $C_n(\mathbf{z}_n)$ does not depend on the state $\mathbf{x}_n$ and is hence typically irrelevant; 
we will ignore it for now and consider it only at the end of Section \ref{sec:approx_calc_JLF}. 
Thus, according to \eqref{eq:joint_likelihood_exp}, for global inference based on the all-sensors measurement vector $\mathbf{z}_n$, 
each sensor needs to know $S_n(\mathbf{z}_n,\mathbf{x}_n)$ as a function of $\mathbf{x}_n$, for the observed (fixed) $\mathbf{z}_n$. 
However, 
calculation of $S_n(\mathbf{z}_n,\mathbf{x}_n)$ at a given sensor
according to \eqref{eq:sum_Sz_exp} presupposes that the sensor knows the measurements $\mathbf{z}_{n,k}$ and
the functions $\mathbf{a}_{n,k}(\cdot)$, $\mathbf{b}_{n,k}(\cdot)$, and $d_{n,k}(\cdot)$ 
of \emph{all} sensors, i.e., for all $k$. 
Transmitting the necessary information from each sensor to each other sensor 
may be infeasible. 

\vspace{-.5mm}

\subsection{Approximation of the Exponential Family}\label{sec:approx_JLF_exp}

\vspace{.8mm}

A powerful 
approach to diffusing local information through a wireless sensor network is given by iterative consensus algorithms, which 
require only communications with 
neighboring sensors and 
are robust to failing communication links and changing network topologies \cite{OlfatiSaber07consensus}.
Unfortunately, a consensus-based distributed calculation of $S_n(\mathbf{z}_n,\mathbf{x}_n)$ 
is not possible in general because the terms of the sum in \eqref{eq:sum_Sz_exp} depend on the unknown state $\mathbf{x}_n$. 
Therefore, we will use an approximation of $S_n(\mathbf{z}_n,\mathbf{x}_n)$ that involves a set of coefficients 
not dependent on $\mathbf{x}_n$. 
This approximation is induced by the following approximations of the functions 
$\mathbf{a}_{n,k}(\mathbf{x}_n)$ and $d_{n,k}(\mathbf{x}_n)$ in terms of given basis functions 
${\{{\varphi}_{n,r}(\mathbf{x}_n)\}}_{r=1}^{R_a}$ 
and ${\{\psi_{n,r}(\mathbf{x}_n)\}}_{r=1}^{R_d}$, 
respectively:
\vspace{-1mm}
\begin{align} 
\mathbf{a}_{n,k}(\mathbf{x}_n) &\,\approx\, \tilde{\mathbf{a}}_{n,k}(\mathbf{x}_n) \,\triangleq\, \sum_{r=1}^{R_a} \bm{\alpha}_{n,k,r} \,{\varphi}_{n,r}(\mathbf{x}_n)
\label{eq:a_approx}\\[1mm]
d_{n,k}(\mathbf{x}_n) &\,\approx\, \tilde{d}_{n,k}(\mathbf{x}_n) \,\triangleq\, \sum_{r=1}^{R_d}\gamma_{n,k,r} \,\psi_{n,r}(\mathbf{x}_n) \,.
\label{eq:d_approx}
\end{align}
Here, $\bm{\alpha}_{n,k,r} \!\in\! \mathbb{R}^{q}$ and $\gamma_{n,k,r} \!\in\! \mathbb{R}$ are expansion 
coefficients that do not depend on $\mathbf{x}_n$. 
(For simplicity, the $\bm{\alpha}_{n,k,r}$ are referred to as coefficients, even though they are vector-valued.) 
The basis functions ${\varphi}_{n,r}(\mathbf{x}_n)$ and $\psi_{n,r}(\mathbf{x}_n)$ do not depend on 
$k$, 
i.e., the same basis functions are used by all sensors. 
They are allowed to depend on 
$n$, even though time-independent basis functions may often be sufficient. 
We assume that sensor $k$ knows the basis functions ${\varphi}_{n,r}(\mathbf{x}_n)$ and $\psi_{n,r}(\mathbf{x}_n)$, 
as well as the coefficients $\bm{\alpha}_{n,k,r}$ and $\gamma_{n,k,r}$ corresponding to its own functions 
$\mathbf{a}_{n,k}(\mathbf{x}_n)$ and $d_{n,k}(\mathbf{x}_n)$, respectively; however, it does not know the coefficients of other sensors, 
$\bm{\alpha}_{n,k'\!,r}$ and $\gamma_{n,k'\!,r}$ with $k' \!\not= k$. 
The coefficients $\bm{\alpha}_{n,k,r}$ and $\gamma_{n,k,r}$ can either be precomputed, or each sensor can calculate them locally. A method for 
calculating these coefficients will be reviewed  
in Section \ref{sec:LS_Approx}.
 
Substituting $\tilde{\mathbf{a}}_{n,k}(\mathbf{x}_n)$ for $\mathbf{a}_{n,k}(\mathbf{x}_n)$ and $\tilde{{d}}_{n,k}(\mathbf{x}_n)$ for ${d}_{n,k}(\mathbf{x}_n)$ 
in \eqref{eq:sum_Sz_exp}, we obtain the following approximation of 
$S_n(\mathbf{z}_n,\mathbf{x}_n)$:
\vspace*{-2mm}
\begin{align}
\tilde{S}_n(\mathbf{z}_n,\mathbf{x}_n)  &\,\triangleq\, \sum_{k=1}^{K}\big[ \tilde{\mathbf{a}}_{n,k}^{\top}(\mathbf{x}_n) \ist \mathbf{b}_{n,k}(\mathbf{z}_{n,k}) 
  - \tilde{d}_{n,k}(\mathbf{x}_n) \big] \label{eq:approx_sum_exp_3_0}\\
&\eq \sum_{k=1}^{K} \Bigg[   
  \Bigg( \sum_{r=1}^{R_a} \bm{\alpha}_{n,k,r}^{\top} \ist {\varphi}_{n,r}(\mathbf{x}_n) \Bigg)  \ist \mathbf{b}_{n,k}(\mathbf{z}_{n,k}) \nonumber\\[-1mm]
&\rule{32mm}{0mm} - \sum_{r=1}^{R_d}\gamma_{n,k,r} \ist \psi_{n,r}(\mathbf{x}_n) \Bigg] \,. \nonumber 
\end{align} 
By changing the order of summation, 
we obtain 
further
\be
\tilde{S}_n(\mathbf{z}_n,\mathbf{x}_n) 
\eq \sum_{r=1}^{R_a}  {A}_{n,r}(\mathbf{z}_n) \ist {\varphi}_{n,r}(\mathbf{x}_n)  \ist-\ist \sum_{r=1}^{R_d} \Gamma_{\rmv n,r} \ist \psi_{n,r}(\mathbf{x}_n) \,, 
\label{eq:approx_sum_exp_4}
\vspace{-1.5mm}
\ee
with
\vspace{-1.5mm}
\be
{A}_{n,r}(\mathbf{z}_n) \rmv\,\triangleq\,\rmv \sum_{k=1}^{K}\bm{\alpha}_{n,k,r}^{\top} \mathbf{b}_{n,k}(\mathbf{z}_{n,k}) \,, \qquad 
\Gamma_{\rmv n,r} \triangleq \sum_{k=1}^{K}\gamma_{n,k,r}\,.
\label{eq:statistic_exp} 
\ee
Finally, substituting $\tilde{S}_{n}(\mathbf{z}_{n},\mathbf{x}_{n})$ from \eqref{eq:approx_sum_exp_4} for 
$S_{n}(\mathbf{z}_{n},\mathbf{x}_{n})$ in \eqref{eq:joint_likelihood_exp}, an approximation of the JLF is obtained  
as
\begin{align}
&\tilde{f}(\mathbf{z}_n|\mathbf{x}_n) \,\propto\, \exp\rmv\rmv \big(\tilde{S}_n(\mathbf{z}_n,\mathbf{x}_n) \big)\nonumber\\[1mm]
  &\rule{5mm}{0mm}\eq \exp\!\Bigg( \sum_{r=1}^{R_a}  {A}_{n,r}(\mathbf{z}_n) \ist {\varphi}_{n,r}(\mathbf{x}_n)  \ist-\ist \sum_{r=1}^{R_d} \Gamma_{\rmv n,r} \ist \psi_{n,r}(\mathbf{x}_n) \Bigg) \,.
\label{eq:approx-JLF-exp}
\end{align}
This shows that a sensor that knows 
${A}_{n,r}(\mathbf{z}_n)$ and $\Gamma_{\rmv n,r}$ can evaluate an approximation of the JLF 
(up to a $\mathbf{z}_n$-dependent but $\mathbf{x}_n$-independent normalization factor) for all values of $\mathbf{x}_n$.
In fact, the vector 
of all coefficients ${A}_{n,r}(\mathbf{z}_n)$ and $\Gamma_{\rmv n,r}$, 
$\tilde{\mathbf{t}}_n(\mathbf{z}_n) \triangleq \big(A_{n,1}(\mathbf{z}_n) \,\cdots\, A_{n,R_a}(\mathbf{z}_n) \;\, \Gamma_{\rmv n,1} \,\cdots\, \Gamma_{\rmv n,R_d}\big)^{\top}\!$, 
can be viewed as a \emph{sufficient statistic} \cite{kay1998fundamentals} that epitomizes the total measurement $\mathbf{z}_n$
within the limits of our approximation.  
Because of expression \eqref{eq:approx-JLF-exp}, this sufficient statistic 
fully describes the approximate JLF $\tilde{f}(\mathbf{z}_n|\mathbf{x}_n)$ 
as a function of $\mathbf{x}_n$.

The expressions \eqref{eq:approx_sum_exp_4} and \eqref{eq:statistic_exp} allow a distributed calculation of $\tilde{S}_{n}(\mathbf{z}_{n},\mathbf{x}_{n})$ and, in turn, 
of $\tilde{f}(\mathbf{z}_n|\mathbf{x}_n)$ by means of consensus algorithms, 
due to the following key facts. 
(i) The coefficients ${A}_{n,r}(\mathbf{z}_{n})$ and $\Gamma_{\rmv n,r}$ do not depend on the state $\mathbf{x}_n$ but contain the information 
of all sensors (the sensor measurements $\mathbf{z}_{n,k}$ and approximation coefficients $\bm{\alpha}_{n,k,r}$ and $\gamma_{n,k,r}$ for all $k$). 
(ii) The state $\mathbf{x}_n$ enters into 
$\tilde{S}_{n}(\mathbf{z}_{n},\mathbf{x}_{n})$ only via the functions ${\varphi}_{n,r}(\cdot)$ and $\psi_{n,r}(\cdot)$, which are sensor-independent and known to each sensor. 
(iii) According to \eqref{eq:statistic_exp}, the coefficients ${A}_{n,r}(\mathbf{z}_{n})$ and $\Gamma_{\rmv n,r}$ are sums in which each term
contains only local information of a single sensor. 
These facts form the basis of the LC method, which will be presented in Section \ref{sec:approx_calc_JLF}. 

Examples of basis functions ${\varphi}_{n,r}(\cdot)$ and $\psi_{n,r}(\cdot)$ are monomials (see the polynomial expansion discussed below), 
orthogonal polynomials, and Fourier basis functions.
The choice of the basis functions affects the accuracy, computational complexity, and communication requirements of the LC 
\vspace{1.5mm}
method.

\noindent {\bf Example---polynomial approximation.\,}
A simple example of a basis expansion approximation \eqref{eq:a_approx} is given by the polynomial approximation 
\be
\tilde{\mathbf{a}}_{n,k}(\mathbf{x}_n) \eq \sum_{\mathbf{r}=\mathbf{0}}^{R_p} \bm{\alpha}_{n,k,\mathbf{r}} \prod_{m=1}^M \rmv x_{n,m}^{r_m} \,, 
\label{eq:polynomial_approx_0} 
\ee
where $\mathbf{r} \!\triangleq\! (r_1 \rmv\cdots\ist r_M) \!\in\! \{0,\ldots,R_p\}^M\rmv$; 
$R_p$ is the degree of the multivariate vector-valued polynomial $\tilde{\mathbf{a}}_{n,k}(\mathbf{x}_n)$; 
$\sum_{\mathbf{r}=\mathbf{0}}^{R_p}$ is short for $\sum_{r_{1}=0}^{R_p}\rmv\cdots\rmv\sum_{r_M=0}^{R_p}$ with the constraint $\sum_{m=1}^M \! r_m \le R_p$;
and $\bm{\alpha}_{n,k,\mathbf{r}} \!\in\! \mathbb{R}^{q}$ is the coefficient vector associated with the basis function (monomial) 
$\varphi_{n,\mathbf{r}}(\mathbf{x}_n) = \prod_{m=1}^M \rmv x_{n,m}^{r_m}$ 
(here, $x_{n,m}$ denotes the $m\ist$th entry of $\mathbf{x}_n$). 
We can rewrite \eqref{eq:polynomial_approx_0} in the form of \eqref{eq:a_approx} by a suitable index mapping
$(r_1 \rmv\cdots\ist r_M) \!\in\! \{0,\ldots,R_p\}^M\rmv \leftrightarrow r \rmv\in\rmv  \{1,\ldots,R_a\}$, where $R_a = {{\ist R_p+M} \choose {\ist R_p}} $. 
An analogous polynomial basis expansion can be used for $\tilde{d}_{n,k}(\mathbf{x}_n)$ in \eqref{eq:d_approx}.
The polynomial basis expansion will be further considered in Section \ref{sec:gauss_pol}.

\vspace{-1.5mm}
 
\subsection{Least Squares Approximation} \label{sec:LS_Approx}

\vspace{.8mm}

A convenient method for calculating the approximations $\tilde{\mathbf{a}}_{n,k}(\mathbf{x}_{n})$ in \eqref{eq:a_approx} and 
$\tilde{d}_{n,k}(\mathbf{x}_{n})$ in \eqref{eq:d_approx} is given by least squares (LS) fitting \cite{bjorck1996numerical, bevington2003data, allaire2008numerical}. 
We first discuss the calculation of the coefficients ${\{ \bm{\alpha}_{n,k,r} \}}_{r=1}^{R_a}$ 
of $\tilde{\mathbf{a}}_{n,k}(\mathbf{x}_{n})$ at time $n$ and sensor $k$.
Consider $J$ data pairs $\big\{ \big(\mathbf{x}_{n,k}^{(j)} \ist, \mathbf{a}_{n,k}({\mathbf{x}}_{n,k}^{(j)}) \big)\big\}_{j=1}^{J}$, where the state points 
${\mathbf{x}}_{n,k}^{(j)}$ are chosen to ``cover'' those regions of the $\mathbf{x}_{n}$ space $\mathbb{R}^M\rmv$ 
where the JLF is expected to be evaluated when estimating $\mathbf{x}_{n}$. 
In particular, in the distributed PF application to be considered in Sections \ref{sec:distributed-particle-filtering} and \ref{sec:distributed-gaussian-particle-filtering}, 
the ${\mathbf{x}}_{n,k}^{(j)}$ will be the predicted particles. 
With LS fitting, the coefficients $\bm{\alpha}_{n,k,r}$ are calculated such that the sum of the squared approximation errors at the state points ${\mathbf{x}}_{n,k}^{(j)}$, 
i.e., $\sum_{j=1}^{J} \rmv\rmv \big\| \tilde{\mathbf{a}}_{n,k}({\mathbf{x}}_{n,k}^{(j)}) - \mathbf{a}_{n,k}({\mathbf{x}}_{n,k}^{(j)}) \big\|^2\rmv$, is minimized.

To describe the solution to this minimization problem, we define the
coefficient matrix $\mathbf{Y}_{n,k} \triangleq \big( \bm{\alpha}_{n,k,1} \ist\cdots \ist \bm{\alpha}_{n,k,R_a} \big)^{\!\top} \!\!\in \mathbb{R}^{R_a \times q}\rmv$,
whose rows are the coefficient vectors ${\{ \bm{\alpha}_{n,k,r} \}}_{r=1}^{R_a}$. Furthermore, 
let
\begin{align*}
\bm{\Phi}_{n,k} &\ist\triangleq\ist \left(\!\! \begin{array}{ccc}
\varphi_{n,1}(\mathbf{x}_{n,k}^{(1)}) \!&\!\! \cdots \!\!&\! \varphi_{n,R_a}(\mathbf{x}_{n,k}^{(1)}) \\[-1mm]
\vdots \!&\!\! \!\!&\! \vdots \\[-.5mm]
\varphi_{n,1}(\mathbf{x}_{n,k}^{(J)}) \!&\!\! \cdots \!\!&\! \varphi_{n,R_a}(\mathbf{x}_{n,k}^{(J)}) \end{array}\!\!\!\right) \!\in \mathbb{R}^{J \times R_a},\\[1mm]
\mathbf{A}_{n,k} &\ist\triangleq\ist \big( \ist\mathbf{a}_{n,k}({\mathbf{x}}_{n,k}^{(1)}) \,\cdots\,  \mathbf{a}_{n,k}({\mathbf{x}}_{n,k}^{(J)}) \ist\big)^{\!\top} 
  \!\rmv\in\mathbb{R}^{J \times q} .
\end{align*}
Then the LS solution for the coefficients ${\{ \bm{\alpha}_{n,k,r} \}}_{r=1}^{R_a}$ is given 
by \cite{bjorck1996numerical}
\[
\mathbf{Y}_{n,k} \eq \big(\bm{\Phi}_{n,k}^{\top}\bm{\Phi}_{n,k} \big)^{-1} \ist \bm{\Phi}_{n,k}^{\top} \ist \mathbf{A}_{n,k} \,. 
\]
Here, we assume that $J \!\ge\! R_a$ and that the columns of $\bm{\Phi}_{n,k}$ are 
linearly independent, so that 
$\bm{\Phi}_{n,k}^{\top}\bm{\Phi}_{n,k}$ is nonsingular. 
Note that $J \!\ge\! R_a$ means that the number of state points $\mathbf{x}_{n,k}^{(j)}$ is not smaller than the number of basis functions ${\varphi}_{n,r}(\mathbf{x}_n)$,
for any given $n$ and $k$.

Similarly, the LS solution for the coefficients ${\{\gamma_{n,k,r}\}}_{r=1}^{R_d}$ of $\tilde{d}_{n,k}(\mathbf{x}_{n})$ in \eqref{eq:d_approx} is obtained 
as $\bm{\gamma}_{n,k} = \big(\bm{\Psi}_{n,k}^{\top}\bm{\Psi}_{n,k} \big)^{-1} \ist \bm{\Psi}_{n,k}^{\top} \mathbf{d}_{n,k}$, 
where $\bm{\gamma}_{n,k} \!\triangleq\! \big( \gamma_{n,k,1} \ist\cdots\gamma_{n,k,R_d} \big)^{\!\top}$\linebreak 
$\in\rmv \mathbb{R}^{R_d}\rmv$,
$\bm{\Psi}_{n,k}\in \mathbb{R}^{J \times R_d}$ is defined like $\bm{\Phi}_{n,k}$ but with ${\{{\varphi}_{n,r}(\cdot)\}}_{r=1}^{R_a}$ replaced by ${\{{\psi}_{n,r}(\cdot)\}}_{r=1}^{R_d}$,  
and $\mathbf{d}_{n,k} \triangleq$\linebreak 
$\big( \ist d_{n,k}({\mathbf{x}}_{n,k}^{(1)}) \,\cdots\,  d_{n,k}({\mathbf{x}}_{n,k}^{(J)}) \ist\big)^{\!\top}\!\rmv\in\mathbb{R}^J\rmv$. 
Here, we assume that $J \!\ge\! R_d$ and that the columns of $\bm{\Psi}_{n,k}$ are 
linearly independent. To summarize, the number of state points $\mathbf{x}_{n,k}^{(j)}$ must satisfy $J \ge \max \{R_a,R_d\}$ for any given $n$ and $k$.

\vspace{-.5mm}

\section{Likelihood Consensus}\label{sec:likelihood-cons_exp}

\vspace{.5mm}

We now present the LC algorithm for local likelihood functions belonging to the exponential family, using the approximation of the JLF discussed in 
Section \ref{sec:approx-exp}. 
Subsequently, we will consider a class of JLFs for which an approximation is not needed.

\vspace{-1.5mm}

\subsection{Distributed Calculation of the Approximate JLF -- The LC Algorithm}\label{sec:approx_calc_JLF}

\vspace{.8mm}

Based on the sum expressions \eqref{eq:statistic_exp}, the sufficient statistic 
$\tilde{\mathbf{t}}_n(\mathbf{z}_n) = \big(A_{n,1}(\mathbf{z}_n) \,\cdots\, A_{n,R_a}(\mathbf{z}_n) \;\, \Gamma_{\rmv n,1} \,\cdots\, \Gamma_{\rmv n,R_d}\big)^{\top}\!$ 
can be computed at each sensor by means of a 
distributed, iterative consensus 
algorithm that requires only communications between neighboring sensors. Here, we use 
a \emph{linear} consensus algorithm \cite{OlfatiSaber07consensus} for simplicity; however, 
other consensus algorithms (e.g., \cite{georgopoulos2009nonlinear}) as well as gossip algorithms (e.g., \cite{dimakis2010gossip})
could be used as well. In what follows, the superscript $^{(i)}$ denotes the iteration index and 
$\mathcal{N}_k \rmv\subseteq\rmv \{1,\ldots,K\} \!\setminus\! \{k\}$ denotes a fixed set of sensors that are neighbors of sensor $k$. 
For simplicity, we only discuss the calculation of ${A}_{n,r}(\mathbf{z}_n)$, since the same principles 
apply to the calculation of $\Gamma_{\rmv n,r}$ in a straightforward manner. We explain the operations performed by a fixed sensor $k$; note that
such operations are performed by all sensors simultaneously.

At time $n$, to compute 
${A}_{n,r}(\mathbf{z}_n) = \sum_{k'=1}^{K} \bm{\alpha}_{n,k'\!,r}^{\top}$\linebreak 
$\times\, \mathbf{b}_{n,k'}(\mathbf{z}_{n,k'})$, sensor $k$ 
first initializes its local ``state'' 
as ${\zeta}_{k}^{(0)} \triangleq \bm{\alpha}_{n,k,r}^{\top} \mathbf{b}_{n,k}(\mathbf{z}_{n,k})$.
This 
involves only the quantities 
$\mathbf{z}_{n,k}$, 
$\bm{\alpha}_{n,k,r}$, 
and 
$\mathbf{b}_{n,k}(\cdot)$, \emph{all of which are available at sensor $k$}; thus, no communication is required at this initialization stage. 
Then, at the $i\ist$th iteration of the consensus algorithm ($i\!\in\!\{1,2,\ldots\}$), the following two steps are performed by sensor $k$:

\begin{itemize}

\item 
Using the previous 
local state ${\zeta}_{k}^{(i-1)}\rmv$ and the previous 
neighbor states ${\zeta}_{k'}^{(i-1)}\rmv$, $k' \!\in\! \mathcal{N}_k$ 
(which were received by sensor $k$ at the previous iteration), the local state of sensor $k$  is updated according to 
\[
{\zeta}_{k}^{(i)} \rmv\eq \omega_{k,k}^{(i)} \ist{\zeta}_{k}^{(i-1)} +\rmv \sum_{k'\in\mathcal{N}_{k}} \!\omega_{k,k'}^{(i)} \ist{\zeta}_{k'}^{(i-1)} .
\vspace{-1mm}
\]
The choice of the weights $\omega_{k,k'}^{(i)}$ is discussed
in 
\cite{xiao2004fast, xiao2005scheme}.
Here, we use the Metropolis weights \cite{xiao2005scheme} 
\vspace{.5mm}
\[
\omega_{k,k'}^{(i)} \equiv \omega_{k,k'} \rmv\eq\rmv
\begin{cases}
\displaystyle\frac{1}{1\ist+\,\max\{|\mathcal{N}_k|,|\mathcal{N}_{k'}|\}} \,,  & k' \!\ne\rmv k \,, \\[2.5mm]
1-\sum_{k''\in \mathcal{N}_k} \omega_{k,k''} \,, & k' \!=\rmv k \,,
\end{cases} 
\vspace{.5mm}
\]
where $|\mathcal{N}_k|$ denotes the number of neighbors of sensor 
$k$. (We note that knowledge at sensor $k$ of $|\mathcal{N}_k|$ and $|\mathcal{N}_{k'}|$, $k' \!\in\! \mathcal{N}_{k}$ 
is not required by certain other choices of the weights \cite{xiao2005scheme}.) 
\vspace{1mm}

\item 
The new local state ${\zeta}_{k}^{(i)}$ is broadcast to all neighbors 
$k' \!\in\! \mathcal{N}_k$.

\vspace{1mm}

\end{itemize}

\noindent These two steps are repeated 
in an iterative manner until a desired degree of convergence is reached. 

If the communication graph of the sensor network is connected, the state ${\zeta}_{k}^{(i)}$ of each sensor $k$ 
converges to the average $\frac{1}{K} \sum_{k'=1}^{K}\bm{\alpha}_{n,k'\!,r}^{\top} \mathbf{b}_{n,k'}(\mathbf{z}_{n,k'}) = \frac{1}{K} \ist {A}_{n,r}(\mathbf{z}_n)$ 
as $i \!\to\! \infty$ \cite{OlfatiSaber07consensus}. Therefore, after 
convergence, the 
states  ${\zeta}_{k}^{(i \to \infty)}$ of all sensors are equal and hence 
a consensus on the value of 
$\frac{1}{K} \ist {A}_{n,r}(\mathbf{z}_n)$ is achieved.
For a finite number $i_{\text{max}}$ of iterations, the 
states ${\zeta}_{k}^{(i_{\text{max}})}$ will be (slightly) different for different sensors $k$
and also 
from the desired value $\frac{1}{K} \ist {A}_{n,r}(\mathbf{z}_n)$.
In what follows, we assume that $i_{\text{max}}$ is sufficiently large so that 
$K{\zeta}_{k}^{(i_{\text{max}})} \approx {A}_{n,r}(\mathbf{z}_n)$ with sufficient accuracy, for all $k$. 
(In the simulations presented in Section \ref{sec:sim-results}, $i_{\text{max}}\in\{7,8,9,10\}$, which arguably does not imply impractical communication requirements.) 
Note that in order to calculate the coefficient ${A}_{n,r}(\mathbf{z}_n)$ 
from ${\zeta}_{k}^{(i_{\text{max}})}$, each sensor needs to know 
$K$.  This information may be provided to each sensor beforehand, 
or some distributed algorithm for counting the number of sensors may be employed (e.g., \cite{mosk2008fast}).

The consensus-based calculations of all 
${A}_{n,r}(\mathbf{z}_n)$, $r=1,\dots,R_a$ and all $\Gamma_{\rmv n,r}$, $r=1,\dots,R_d$ are executed simultaneously, 
and their iterations are synchronized. These consensus algorithms taken together 
form the LC algorithm, which is stated in what 
\vspace{5mm}
follows. 

{
\hrule
\begin{center}\vspace{-1.3mm}
{\small\sc Algorithm 1:\, Likelihood Consensus (LC)}\vspace{-1.2mm}

\end{center}
\hrule
\vspace{3mm}

\small
\renewcommand{\baselinestretch}{1.1}\normalsize\small

\noindent
At time $n$, the following steps 
are performed by sensor $k$ (analogous steps 
are performed by all sensors simultaneously).

\vspace{1.5mm}

\begin{enumerate}

\item Calculate the coefficients ${\{ \bm{\alpha}_{n,k,r} \}}_{r=1}^{R_a}$ and ${\{ \gamma_{n,k,r} \}}_{r=1}^{R_d}$ 
of the approximations \eqref{eq:a_approx} and \eqref{eq:d_approx}. 

\vspace{2mm}

\item \emph{Consensus algorithm---${A}_{n,r}(\mathbf{z}_n)$}: For each $r = 1,\dots,R_a$: 

\vspace{.5mm}

\begin{enumerate}

\item 
Initialize the local state 
as ${\zeta}_k^{(0)} \!= \bm{\alpha}_{n,k,r}^{\top} \mathbf{b}_{n,k}(\mathbf{z}_{n,k})$.

\vspace{1.5mm}

\item For $i=1,2,\dots,i_{\text{max}}$ (here, $i_{\text{max}}$ is a predetermined iteration count or determined by the condition that 
$\big| {\zeta}_{k}^{(i)} \!- {\zeta}_{k}^{(i-1)} \big|$ falls below a given threshold):

\begin{itemize}

\vspace{.3mm}

\item Update the local state according to ${\zeta}_{k}^{(i)} = \omega_{k,k}^{(i)} \ist {\zeta}_{k}^{(i-1)}$\linebreak 
$+\ist \sum_{k'\in\mathcal{N}_{k}} \!\omega_{k,k'}^{(i)} \ist{\zeta}_{k'}^{(i-1)}$.

\vspace{.2mm}

\item Broadcast the new state ${\zeta}_{k}^{(i)}$ to all neighbors $k' \!\rmv\in\rmv \mathcal{N}_k$.

\vspace{1mm}

\end{itemize}

\item Calculate $\tilde{A}_{n,r}(\mathbf{z}_n) \triangleq K{\zeta}_{k}^{(i_{\text{max}})}$.

\end{enumerate}

\vspace{1mm}

\item \emph{Consensus algorithm---$\Gamma_{\rmv n,r}$}: For each $r = 1,\dots,R_d$:

\vspace{.5mm}

\begin{enumerate}

\item 
Initialize the local state 
as ${\zeta}_k^{(0)} \!= \gamma_{n,k,r}$.

\vspace{1mm}

\item Same as 2b).

\vspace{1mm}

\item Calculate $\tilde{\Gamma}_{\rmv n,r} \triangleq K{\zeta}_{k}^{(i_{\text{max}})}$.

\end{enumerate}

\vspace{.7mm}

\end{enumerate}

Finally, by substituting $\tilde{A}_{n,r}(\mathbf{z}_n)$ for ${A}_{n,r}(\mathbf{z}_n)$ and $\tilde{\Gamma}_{\rmv n,r}$ for $\Gamma_{\rmv n,r}$ in \eqref{eq:approx-JLF-exp},
sensor $k$ is able to obtain a consensus approximation of the approximate JLF $\tilde{f}(\mathbf{z}_n|\mathbf{x}_n)$ 
for any given value of $\mathbf{x}_n$.
}
\vspace{2mm}
\hrule
\vspace{3mm}

Because one consensus algorithm has to be executed for each ${A}_{n,r}(\mathbf{z}_n)$, $r = 1,\dots,R_a$ 
and 
$\Gamma_{\rmv n,r}$, $r = 1,\dots,R_d\ist$, the number of consensus algorithms that are executed 
simultaneously 
is $N_c = R_a + R_d$. This is also the number of real numbers broadcast by each sensor in each iteration of the LC algorithm. 
It is important to note that $R_a$ and $R_d$ do not depend on the dimensions $N_{n,k}$ of the 
measurement vectors $\mathbf{z}_{n,k}$, and thus the communication requirements of LC do not depend on the $N_{n,k}$.  
This is particularly advantageous in the case of high-dimensional 
measurements. 
However, $R_a$ and $R_d$ usually grow with the dimension $M$ of the state vector $\mathbf{x}_n$. In particular, 
if the $M$D basis ${\{ {\varphi}_{n,r}(\mathbf{x}_n) \}}_{r=1}^{R_a}$ is constructed as the $M$-fold tensor product of a 1D basis 
${\{ \tilde{\varphi}_{n,\tilde{r}}(x) \}}_{\tilde{r}=1}^{\tilde{R}_a}$, then $R_a = \tilde{R}_a^M$, and similarly for the $M$D basis 
${\{ {\psi}_{n,r}(\mathbf{x}_n) \}}_{r=1}^{R_d}$.

So far, we have disregarded the normalization factor $C_n(\mathbf{z}_n)$ occurring in \eqref{eq:joint_likelihood_exp}.
If this factor is required at each sensor, it can also be computed by a consensus algorithm. 
\vspace{-2mm}
From \eqref{eq:joint_likelihood_exp_C},
\[
\log C_n(\mathbf{z}_n) \eq \sum_{k=1}^{K} \log c_{n,k}(\mathbf{z}_{n,k}) \,.
\]
Since this is a sum and $c_{n,k}(\mathbf{z}_{n,k})$ is known to each sensor, a consensus algorithm can again be used
for a distributed calculation of $\log C_n(\mathbf{z}_n)$.

\vspace{-1.2mm}

\subsection{Distributed Calculation of the Exact JLF} \label{sec:LC-exact}

\vspace{.8mm}

The basis expansion approximations \eqref{eq:a_approx} and \eqref{eq:d_approx} 
can be avoided if the JLF $f(\mathbf{z}_n|\mathbf{x}_n)$ has a special structure.
In that case, the \emph{exact} JLF can be computed in a distributed way, up to errors that are only due to the limited number of consensus iterations
performed. We note that the special structure considered now is compatible with the exponential family structure
considered so far, but it does not presuppose that structure. 

Let $\mathbf{t}_n(\mathbf{z}_n) = \big( \ist t_{n,1}(\mathbf{z}_n) \,\cdots\, t_{n,P}(\mathbf{z}_n) \big)^{\!\top}\rmv$ be a sufficient statistic
for the estimation problem corresponding to $f(\mathbf{z}_n|\mathbf{x}_n)$. 
According to the Neyman-Fisher factorization theorem \cite{kay1998fundamentals}, $f(\mathbf{z}_n|\mathbf{x}_n)$ can then be written 
as 
\be
f(\mathbf{z}_n|\mathbf{x}_n) \ist=\ist f_{1}(\mathbf{z}_n) \ist\ist f_{2} \big(\mathbf{t}_n(\mathbf{z}_n),\mathbf{x}_n\big) \,.
\label{eq:f_suff_statistic}
\ee
Typically, the factor $f_{1}(\mathbf{z}_n)$ can be disregarded since it does not depend on $\mathbf{x}_n$. 
Thus, $\mathbf{t}_n(\mathbf{z}_n)$ epitomizes the total measurement $\mathbf{z}_n$, in 
that a sensor that knows 
$\mathbf{t}_n(\mathbf{z}_n)$ and 
$f_{2}(\cdot \ist,\cdot)$ is able to evaluate 
the JLF $f(\mathbf{z}_n|\mathbf{x}_n)$ (up to an irrelevant factor) for any value of $\mathbf{x}_n$.
Suppose further that the components of $\mathbf{t}_n(\mathbf{z}_n)$ have the 
form
\be
t_{n,p}(\mathbf{z}_n) \eq\rmv \sum_{k=1}^{K} \eta_{n,k,p}(\mathbf{z}_{n,k}) \,, \quad\;\; p = 1,\ldots,P \,,
\label{eq:exact_suff_statistic}
\ee
with arbitrary functions $\eta_{n,k,p}(\cdot)$, and that sensor $k$ knows its own functions $\eta_{n,k,p}(\cdot)$ 
but not 
$\eta_{n,k'\!,p}(\cdot)$, $k' \!\rmv\not=\! k$.
Based on the sum expression \eqref{eq:exact_suff_statistic}, we can then use consensus algorithms as described in Section \ref{sec:approx_calc_JLF}, 
with obvious modifications, to calculate $\mathbf{t}_n(\mathbf{z}_n)$ and, thus, the JLF $f(\mathbf{z}_n|\mathbf{x}_n)$ in a distributed manner.

Clearly, an example where exact calculation of the JLF is possible is the case where $f(\mathbf{z}_n|\mathbf{x}_n)$ belongs to the exponential family 
\eqref{eq:joint_likelihood_exp_0}, with functions $\mathbf{a}_{n,k}(\mathbf{x}_n)$ and ${d}_{n,k}(\mathbf{x}_n)$ that 
can be exactly represented using expansions of the form \eqref{eq:a_approx} and \eqref{eq:d_approx}, i.e.,
$\mathbf{a}_{n,k}(\mathbf{x}_n) = \sum_{r=1}^{R_a} \bm{\alpha}_{n,k,r} \ist {\varphi}_{n,r}(\mathbf{x}_n)$
and $d_{n,k}(\mathbf{x}_n) = \sum_{r=1}^{R_d}\gamma_{n,k,r} \ist \psi_{n,r}(\mathbf{x}_n)$.
This is a special case of \eqref{eq:f_suff_statistic} and \eqref{eq:exact_suff_statistic}, with (cf.\ 
\eqref{eq:approx-JLF-exp})
\[
f_{2}(\mathbf{t}_n(\mathbf{z}_n),\mathbf{x}_n) \ist=\, \exp\!\Bigg( \sum_{p=1}^{P}  t_{n,p}(\mathbf{z}_n) \ist {\rho}_{n,p}(\mathbf{x}_n) \Bigg) \ist ,
\]
where $P = R_a + R_d$ 
\vspace{1mm}
and
\begin{align*}
t_{n,p}(\mathbf{z}_n) &\,=\ist 
\begin{cases}
&\hspace{-3mm}A_{n,p}(\mathbf{z}_n) = \sum_{k=1}^{K}\bm{\alpha}_{n,k,p}^{\top} \mathbf{b}_{n,k}(\mathbf{z}_{n,k}),\\
& \rule{34mm}{0mm}p=1,\dots,R_a, \\[1mm]
&\hspace{-3mm}-\Gamma_{\rmv n,p \ist-R_a} = - \sum_{k=1}^{K}\gamma_{n,k,p \ist-R_a}, \\
& \rule{34mm}{0mm}p=R_a \!+\! 1,\dots,P \ist ,
\end{cases}
\\[1mm]
{\rho}_{n,p}(\mathbf{x}_n) &\,=\ist 
\begin{cases}
{\varphi}_{n,p}(\mathbf{x}_n), & p=1,\dots,R_a,\\[.7mm]
\psi_{n,p \ist-R_a}(\mathbf{x}_n), & p=R_a \!+\! 1,\dots,P \ist .
\end{cases}
\end{align*}
Equivalently, $t_{n,p}(\mathbf{z}_n)$ is of the form \eqref{eq:exact_suff_statistic}, with  
(cf.\ \eqref{eq:statistic_exp})
\[
\eta_{n,k,p}(\mathbf{z}_{n,k}) \,=\ist 
\begin{cases}
\bm{\alpha}_{n,k,p}^{\top} \mathbf{b}_{n,k}(\mathbf{z}_{n,k}), & p=1,\dots,R_a, \\[.5mm]
- \ist \gamma_{n,k,p \ist-R_a}, & p=R_a \!+\! 1,\dots,P \ist .
\end{cases}
\vspace{1mm}
\]

\section{Special Case: Gaussian Measurement Noise}\label{sec:gauss} 

\vspace{.5mm}

In this section, we consider the important special case of (generally nonlinear) measurement functions and
independent additive Gaussian measurement noises at the various sensors. 
We will also develop the application of the polynomial approximation that was briefly introduced in Section \ref{sec:approx_JLF_exp}. 

\vspace{-.7mm}

\subsection{Measurement Model} \label{sec:gauss_model}

\vspace{.8mm}

The dependence of the sensor measurements $\mathbf{z}_{n,k}$ on the state $\mathbf{x}_n$ is described by the local likelihood 
functions $f(\mathbf{z}_{n,k}|\mathbf{x}_n)$. Let us now assume, more specifically, 
that the measurements are modeled as 
\be
\mathbf{z}_{n,k} \eq \mathbf{h}_{n,k}(\mathbf{x}_n)+\mathbf{v}_{n,k} \,, \quad\;\; k = 1,\ldots,K \,,
\label{eq:scalar_meas_model_additive_gauss}
\ee 
where $\mathbf{h}_{n,k}(\cdot)$ is the \emph{measurement function} of sensor $k$ and 
$\mathbf{v}_{n,k} \rmv\sim\rmv \mathcal{N}(\mathbf{0},\mathbf{Q}_{n,k})$ is zero-mean Gaussian measurement noise
that is independent of $\mathbf{x}_{n'}$ for all $n'\rmv$.
We furthermore assume that $\mathbf{v}_{n,k}$ and $\mathbf{v}_{n'\!,k'}$ are independent unless $(n,k) = (n'\!,k')$. 
Under these assumptions, the $\mathbf{z}_{n,k}$ are conditionally independent given $\mathbf{x}_n$, i.e., \eqref{eq:joint_likelihood_prod} holds.
The local likelihood function of sensor $k$ is here given 
by
\begin{align}
&f(\mathbf{z}_{n,k}|\mathbf{x}_n) \nonumber\\[0mm]
&\rule{.1mm}{0mm}\eq \bar{c}_{n,k} \exp\!\bigg(\!\!-\rmv \frac{1}{2} \ist [\mathbf{z}_{n,k} \!-\rmv \mathbf{h}_{n,k}(\mathbf{x}_n)]^\top 
  \mathbf{Q}_{n,k}^{-1} \ist [\mathbf{z}_{n,k} \!-\rmv \mathbf{h}_{n,k}(\mathbf{x}_n)] \rmv \bigg) \, , 
\label{eq:local_likelihood_additive_gauss}
\end{align}
with $\bar{c}_{n,k} \triangleq [(2\pi)^{N_{n,k}}\det\{ \mathbf{Q}_{n,k} \}]^{-1/2}$. Furthermore, using \eqref{eq:joint_likelihood_prod}, the JLF is obtained 
as
\begin{align}
&f(\mathbf{z}_{n}|\mathbf{x}_n) \nonumber\\[0mm]
&\rule{.01mm}{0mm}\eq \bar{c}_{n} \exp\!\Bigg(\!\!-\rmv \frac{1}{2} \ist \sum_{k=1}^K [\mathbf{z}_{n,k} \!-\rmv \mathbf{h}_{n,k}(\mathbf{x}_n)]^\top 
  \mathbf{Q}_{n,k}^{-1} \ist [\mathbf{z}_{n,k} \!-\rmv \mathbf{h}_{n,k}(\mathbf{x}_n)] \rmv \Bigg)  , 
\label{eq:JLF_additive_gauss}\\[-8mm]
&\nonumber
\end{align}
with $\bar{c}_{n} = \prod_{k=1}^K \bar{c}_{n,k}$.

The local likelihood function $f(\mathbf{z}_{n,k}|\mathbf{x}_n)$ in \eqref{eq:local_likelihood_additive_gauss} 
is a special case of the exponential family \eqref{eq:loc_likelihood_exp}, with
\begin{align}
\mathbf{a}_{n,k}(\mathbf{x}_{n})  &\eq \mathbf{h}_{n,k}(\mathbf{x}_n) \,, \label{eq:a_gauss}\\
\mathbf{b}_{n,k}(\mathbf{z}_{n,k})  &\eq  \mathbf{Q}_{n,k}^{-1} \ist \mathbf{z}_{n,k} \,, \nonumber \\[0mm]
c_{n,k}(\mathbf{z}_{n,k})  &\eq \bar{c}_{n,k} \exp\!\bigg(\!\!-\rmv \frac{1}{2}\ist \mathbf{z}_{n,k}^{\top} \mathbf{Q}_{n,k}^{-1} \ist \mathbf{z}_{n,k}\rmv \bigg) \, , \nonumber \\[0mm]
d_{n,k}(\mathbf{x}_{n})  &\eq \frac{1}{2} \ist \mathbf{h}_{n,k}^{\top}(\mathbf{x}_n) \ist \mathbf{Q}_{n,k}^{-1} \ist \mathbf{h}_{n,k}(\mathbf{x}_n) 
 \,. \label{eq:d_gauss}
\end{align}
Consequently 
(see \eqref{eq:sum_Sz_exp}),
\be
S_n(\mathbf{z}_n,\mathbf{x}_n) \eq\rmv \sum_{k=1}^{K} \mathbf{h}_{n,k}^{\top}(\mathbf{x}_n) \ist \mathbf{Q}_{n,k}^{-1} \ist 
  \bigg[\mathbf{z}_{n,k} - \frac{1}{2} \ist \mathbf{h}_{n,k}(\mathbf{x}_n) \bigg] \,.
\label{eq:sum_Sz_exp_gauss}
\ee 
We now approximate $\mathbf{a}_{n,k}(\mathbf{x}_{n})$ and $d_{n,k}(\mathbf{x}_n)$
by truncated basis expansions $\tilde{\mathbf{a}}_{n,k}(\mathbf{x}_n)$ and $\tilde{d}_{n,k}(\mathbf{x}_n)$  
of the form \eqref{eq:a_approx} and \eqref{eq:d_approx}, respectively. 
According to \eqref{eq:a_gauss}, approximating $\mathbf{a}_{n,k}(\mathbf{x}_{n})$ is equivalent to approximating
the sensor measurement function $\mathbf{h}_{n,k}(\mathbf{x}_n)$ (which is also the mean of $f(\mathbf{z}_{n,k}|\mathbf{x}_n)$ 
in \eqref{eq:local_likelihood_additive_gauss}). Thus, 
\vspace{-2mm}
\be
\tilde{\mathbf{a}}_{n,k}(\mathbf{x}_n) \eq \tilde{\mathbf{h}}_{n,k}(\mathbf{x}_n) \eq \sum_{r=1}^{R_a} \bm{\alpha}_{n,k,r} \,{\varphi}_{n,r}(\mathbf{x}_n) \,.
\label{eq:a_approx_h}
\ee
Furthermore, an approximation of $d_{n,k}(\mathbf{x}_n)$ of the form \eqref{eq:d_approx} can be obtained in an
indirect way by substituting in \eqref{eq:d_gauss} the above approximation $\tilde{\mathbf{h}}_{n,k}(\mathbf{x}_n)$ 
for $\mathbf{h}_{n,k}(\mathbf{x}_n)$; this 
yields
\begin{align}
\tilde{d}_{n,k}(\mathbf{x}_{n}) &\eq \frac{1}{2} \ist \tilde{\mathbf{h}}_{n,k}^{\top}(\mathbf{x}_n) \ist \mathbf{Q}_{n,k}^{-1} \ist \tilde{\mathbf{h}}_{n,k}(\mathbf{x}_n)
\label{eq:d_approx_gauss} \\[.5mm]
&\eq \frac{1}{2}  \sum_{r_1=1}^{R_a} \sum_{r_2=1}^{R_a}  \bm{\alpha}_{n,k,r_1}^{\top} \! \mathbf{Q}_{n,k}^{-1}  \ist \bm{\alpha}_{n,k,r_2} \ist 
  {\varphi}_{n,r_1}(\mathbf{x}_n) \ist {\varphi}_{n,r_2}(\mathbf{x}_n) \,.
\label{eq:d_approx_gauss_1} 
\end{align}
Using a suitable index mapping $(r_1,r_2) \rmv\in\rmv \{1,\ldots,R_a\} \!\times\! \{1,\ldots,R_a\} \leftrightarrow r \rmv\in\rmv \{1,\ldots,R_d\}$,
we can write \eqref{eq:d_approx_gauss_1} in the 
form 
\vspace{-2mm}
\eqref{eq:d_approx}: 
\[
\tilde{d}_{n,k}(\mathbf{x}_n) \eq \sum_{r=1}^{R_d}\gamma_{n,k,r} \,\psi_{n,r}(\mathbf{x}_n) \,,
\]
with  $R_d = R_a^2$, $\gamma_{n,k,r} = \frac{1}{2} \ist \bm{\alpha}_{n,k,r_1}^{\top} \! \mathbf{Q}_{n,k}^{-1}  \ist \bm{\alpha}_{n,k,r_2}$, and 
$\psi_{n,r}(\mathbf{x}_n)$\linebreak 
$ = {\varphi}_{n,r_1}(\mathbf{x}_n) \ist {\varphi}_{n,r_2}(\mathbf{x}_n)$. 
It is 
easily verified that with this special basis expansion approximation of $d_{n,k}(\mathbf{x}_n)$, the resulting approximate JLF can be written as
\begin{align*}
&\tilde{f}(\mathbf{z}_{n}|\mathbf{x}_n) \nonumber\\[0mm]
&\rule{-1mm}{0mm}=\ist \bar{c}_{n} \exp\!\Bigg(\!\!-\rmv \frac{1}{2} \ist \sum_{k=1}^K [\mathbf{z}_{n,k} \!-\rmv \tilde{\mathbf{h}}_{n,k}(\mathbf{x}_n)]^\top 
  \mathbf{Q}_{n,k}^{-1} \ist [\mathbf{z}_{n,k} \!-\rmv \tilde{\mathbf{h}}_{n,k}(\mathbf{x}_n)] \Bigg) \ist , 
\end{align*}
which is \eqref{eq:JLF_additive_gauss} with $\mathbf{h}_{n,k}(\mathbf{x}_n)$ replaced by $\tilde{\mathbf{h}}_{n,k}(\mathbf{x}_n)$. This means that only the mean
of $f(\mathbf{z}_{n}|\mathbf{x}_n)$ is changed by this approximation. 

In the additive Gaussian noise setting considered, the LC method operates almost 
as in the general case. 
The only difference is in Step 
1 of Algorithm 1: instead of calculating the coefficients $\gamma_{n,k,r}$ directly, using, e.g.,\ a separate 
LS fitting, we obtain them in an indirect way as described above. Hence, the computational complexity is reduced. Note that in general, the directly and indirectly obtained coefficients $\gamma_{n,k,r}$ will be different. 
Furthermore, if the indirectly obtained coefficients are used, the approximate JLF $\tilde{f}(\mathbf{z}_{n}|\mathbf{x}_n)$ 
is a valid pdf in the sense that $\int \tilde{f}(\mathbf{z}_{n}|\mathbf{x}_n) \, d\mathbf{z}_{n} = 1$ holds exactly, not only 
approximately. The number of consensus algorithms that are executed is $N_c = R_a + R_d = R_a + R_a^2$. Again, this does not depend on the dimensions $N_{n,k}$ of the measurement vectors $\mathbf{z}_{n,k}$ since $R_a$ does not depend on $N_{n,k}$. 

\vspace{-1.2mm}

\subsection{Polynomial Approximation}\label{sec:gauss_pol}

\vspace{.8mm}

The polynomial approximation was introduced in Section \ref{sec:approx_JLF_exp}. 
We will now apply it to the case of Gaussian measurement noise studied above. Using \eqref{eq:polynomial_approx_0}, we obtain 
for \eqref{eq:a_approx_h} 
\be
\tilde{\mathbf{a}}_{n,k}(\mathbf{x}_n) \eq \tilde{\mathbf{h}}_{n,k}(\mathbf{x}_n) 
  \eq \sum_{\mathbf{r}=\mathbf{0}}^{R_p} \bm{\alpha}_{n,k,\mathbf{r}} \rmv \prod_{m=1}^M \rmv x_{n,m}^{r_m} \,. 
\label{eq:polynomial_approx}
\ee
Inserting this into \eqref{eq:d_approx_gauss} yields
\be
\tilde{d}_{n,k}(\mathbf{x}_n) \eq \sum_{\mathbf{r}=\mathbf{0}}^{2R_p} \gamma_{n,k,\mathbf{r}} \rmv \prod_{m=1}^M \rmv x_{n,m}^{r_m} \,,
\label{eq:d_2tilde_poly}
\ee
with 
\be
\gamma_{n,k,\mathbf{r}} \eq \frac{1}{2} \, \underset{\mathbf{r}'+\mathbf{r}'' \ist=\ist \mathbf{r}}{\sum_{\mathbf{r}'=\mathbf{0}}^{R_p} 
  \sum_{\mathbf{r}''=\mathbf{0}}^{R_p}} \, \bm{\alpha}_{n,k,\mathbf{r}'}^{\top} \ist \mathbf{Q}_{n,k}^{-1}  \ist \bm{\alpha}_{n,k,\mathbf{r}''} \,.
\label{eq:gamma_exp}
\ee
Next, inserting expressions \eqref{eq:polynomial_approx} and \eqref{eq:d_2tilde_poly} into \eqref{eq:approx_sum_exp_3_0}, we obtain
\be
\tilde{S}_n(\mathbf{z}_n,\mathbf{x}_n)  \eq 
\sum_{k=1}^{K}\sum_{\mathbf{r}=\mathbf{0}}^{2R_p} \beta_{n,k,\mathbf{r}}(\mathbf{z}_{n,k}) \rmv\prod_{m=1}^M \rmv x_{n,m}^{r_m} \,,
\label{eq:approx_sum_exp_3_poly}
\vspace{-1.7mm}
\ee
with 
\be
\hspace*{-1.5mm}\beta_{n,k,\mathbf{r}}(\mathbf{z}_{n,k}) \eq 
\begin{cases}
\bm{\alpha}_{n,k,\mathbf{r}}^{\top} \ist \mathbf{b}_{n,k}(\mathbf{z}_{n,k}) - \gamma_{n,k,\mathbf{r}} \ist, & \mathbf{r} \!\in\! \mathcal{R}_1 \\
- \gamma_{n,k,\mathbf{r}} \ist, & \mathbf{r} \!\in\! \mathcal{R}_2 \,,
\end{cases}
\label{eq:beta_exp}
\ee
where $\mathcal{R}_1$ is the set of all $\mathbf{r} = (r_1 \rmv\cdots\ist r_M) \!\in\! \{0,\ldots,R_p\}^M\rmv$ such that $\sum_{m=1}^M \! r_m \le R_p$ 
and $\mathcal{R}_2$ is the set of all $\mathbf{r} \!\in\! \{0,\ldots,2R_p\}^M \setminus \mathcal{R}_1$ such that $\sum_{m=1}^M \! r_m \le 2R_p$. 
Finally, changing the order of summation in \eqref{eq:approx_sum_exp_3_poly} 
gives 
\be
\tilde{S}_n(\mathbf{z}_n,\mathbf{x}_n) \eq \sum_{\mathbf{r}=\mathbf{0}}^{2R_p} B_{n,\mathbf{r}}(\mathbf{z}_n) \rmv\prod_{m=1}^M \rmv x_{n,m}^{r_m} \,, 
\label{eq:approx_sum_exp_4_poly}
\vspace{-1mm}
\ee
with 
\vspace{1mm}
\be
B_{n,\mathbf{r}}(\mathbf{z}_n) \eq \sum_{k=1}^{K} \beta_{n,k,\mathbf{r}}(\mathbf{z}_{n,k}) \,.
\label{eq:statistic_poly}
\ee
It should be noted that \eqref{eq:approx_sum_exp_4_poly} is a special case of \eqref{eq:approx_sum_exp_4}.
The coefficients $B_{n,\mathbf{r}}(\mathbf{z}_n)$ 
can again be calculated using a consensus algorithm. 
For each time $n$, the number of coefficients $B_{n,\mathbf{r}}(\mathbf{z}_n)$, and hence the number of consensus algorithms that have to be 
executed in parallel, is given by  $N_c = {{\ist 2R_p+M} \choose {\ist 2R_p}} - 1$. Here, the subtraction of $1$ is due to the fact that 
the coefficient $B_{n,\mathbf{r = \mathbf{0}}}(\mathbf{z}_n)$ need not be calculated: according to
\eqref{eq:approx_sum_exp_4_poly}, $B_{n,\mathbf{\mathbf{0}}}(\mathbf{z}_n)$ corresponds to a JLF factor that does not depend on $\mathbf{x}_n$
and is hence irrelevant.

\section{Distributed Particle Filtering}\label{sec:distributed-particle-filtering}

\vspace{.5mm}

In this section, we show how the LC method can be applied to obtain a distributed PF. By way of preparation, 
we first review a standard centralized PF \cite{doucet2001sequential,arulampalam2002tpf,ristic2004bkf}.

\vspace{-1.1mm}

\subsection{Review of Centralized Particle Filtering}\label{subsec:CPF}

\vspace{.8mm}

The centralized PF is implemented at a fusion center that 
knows the all-sensors measurement vector $\mathbf{z}_n$
and the functional form of the JLF $f(\mathbf{z}_n|\mathbf{x}_n)$. The PF maintains a set of samples (or particles) $\big\{ \mathbf{x}_n^{(j)} \big\}_{j=1}^J$ 
and associated weights $\big\{ w_n^{(j)} \big\}_{j=1}^J$, which 
establish the following approximative sample representation of the posterior pdf 
$f(\mathbf{x}_n|\mathbf{z}_{1:n})$: 
\[
f_\delta(\mathbf{x}_n|\mathbf{z}_{1:n}) \,\triangleq\, \sum_{j=1}^{J} w_n^{(j)} \ist \delta \big(\mathbf{x}_n \!- \mathbf{x}_n^{(j)} \big) \,. 
\]
The MMSE estimate in \eqref{eq:mmse_est} can then be approximated by the mean of $f_\delta(\mathbf{x}_n|\mathbf{z}_{1:n})$,
which is equivalent to a weighted sample 
mean:
\be
\hat{\mathbf{x}}_{n} \ist\triangleq \int \rmv \mathbf{x}_n  \, f_\delta(\mathbf{x}_n|\mathbf{z}_{1:n}) \, d\mathbf{x}_n 
\rmv\eq\rmv \sum_{j=1}^{J} w_{n}^{(j)} \ist \mathbf{x}_{n}^{(j)} \,.
\label{eq:pfEstimate_CPF}
\ee 
At each time step $n$, when the 
new measurement vector $\mathbf{z}_n$ becomes available, new particles and weights are calculated
by a PF algorithm that is based on the recursion \eqref{eq:sequ_post_update}. 

Many PF algorithms have been proposed 
\cite{gordon1993novel,doucet2001sequential,arulampalam2002tpf,ristic2004bkf}. Here, 
we 
consider a sequential importance resampling 
filter \cite{gordon1993novel,arulampalam2002tpf}, 
which performs the following steps. 
For initialization ($n \!=\! 0$), 
$J$ particles $\mathbf{x}_0^{(j)}$ are sampled from a prior distribution $f(\mathbf{x}_{0})$,
and the weights are set to $w_0^{(j)} \!\equiv\rmv 1/J$.
Then, 
three steps---resampling, sampling, and weight update---are repeated for every 
$n$. 
In the \emph{resampling step}, $J$ resampled particles $\bar{\mathbf{x}}_{n-1}^{(j)}$ are obtained
by sampling with replacement from the set of previous particles $\big\{{\mathbf{x}}_{n-1}^{(j')}\big\}_{j'=1}^{J}$, where the probability of sampling ${\mathbf{x}}_{n-1}^{(j')}$ 
is $w_{n-1}^{(j')}$. 
In the \emph{sampling step}, for each resampled particle $\bar{\mathbf{x}}_{n-1}^{(j)}$, 
a new, ``predicted'' particle $\mathbf{x}_{n}^{(j)}$ is 
sampled from $f(\mathbf{x}_{n}|\bar{\mathbf{x}}_{n-1}^{(j)})$,
i.e., from the state-transition pdf $f(\mathbf{x}_{n}|\mathbf{x}_{n-1})$ evaluated at $\mathbf{x}_{n-1} = \bar{\mathbf{x}}_{n-1}^{(j)}$.
In the \emph{weight update step},
the weight associated with each particle $\mathbf{x}_n^{(j)}$ is calculated 
\vspace{-.5mm}
as 
\be
w_{n}^{(j)} \rmv\eq \frac{f(\mathbf{z}_{n}|\mathbf{x}_{n}^{(j)})}{\sum_{j'=1}^{J}f(\mathbf{z}_{n}|\mathbf{x}_{n}^{(j')})} \,.
\label{eq:weightUpdate_centr}
\ee 
Finally, the state estimate $\hat{\mathbf{x}}_{n}$ is 
calculated from $\big\{ \big( \mathbf{x}_n^{(j)},$\linebreak 
$w_{n}^{(j)} \big) \big\}_{j=1}^J$ according to \eqref{eq:pfEstimate_CPF}.

\vspace{-.8mm}

\subsection{Distributed Particle Filtering Using LC} \label{subsec:distributed-pf}

\vspace{.8mm}

Next, we 
develop a distributed implementation of the sequential importance resampling  
filter reviewed above, in which 
each sensor acts similarly to 
the fusion center of the centralized PF. 
More specifically, sensor $k$ 
tracks a particle representation of the global posterior $f(\mathbf{x}_n|\mathbf{z}_{1:n})$ using a \emph{local PF}. 
For each $n$, it obtains a state estimate $\hat{\mathbf{x}}_{n,k}$ that is based on $\mathbf{z}_{1:n}$, i.e., the 
past and current measurements of \emph{all} sensors. This requires each sensor to know the JLF $f(\mathbf{z}_{n}|\mathbf{x}_{n})$ 
as a function of the state $\mathbf{x}_{n}$, because the weight update in \eqref{eq:weightUpdate_centr} requires the pointwise evaluation
of the JLF. Therefore, an approximation of the JLF is provided to each sensor in a distributed way by means of 
the LC method. 
No routing of measurements or other sensor-local data is needed; each sensor merely broadcasts information 
to neighboring sensors. 
The algorithm is 
\vspace{4mm}
stated as follows. 

{
{\hrule
\begin{center}\vspace{-1mm}
{\small\sc Algorithm 2:\, LC-based Distributed PF (LC-DPF)}\vspace{-1.2mm}
\end{center}
\hrule}
\vspace{2.8mm}

\small
\renewcommand{\baselinestretch}{1.18}\normalsize\small

\noindent
At 
time $n$, the local PF at sensor $k$ performs the following steps, 
which are identical for all $k$. 
(Note that these steps are essentially analogous to those 
of the centralized PF of Section \ref{subsec:CPF}, except that 
an approximation of the JLF is used.) 

\vspace{1mm}

\begin{enumerate}

\item At the previous time $n\!-\!1$, sensor $k$ calculated
$J$ particles $\mathbf{x}_{n-1,k}^{(j)}$ and 
weights $w_{n-1,k}^{(j)}$, 
which together represent the previous global posterior $f(\mathbf{x}_{n-1}|\mathbf{z}_{1:n-1})$. The first step at time $n$ is a resampling of
$\big\{ \big( \mathbf{x}_{n-1,k}^{(j)},w_{n-1,k}^{(j)} \big) \big\}_{j=1}^{J}$, which produces $J$ resampled particles 
$\bar{\mathbf{x}}_{n-1,k}^{(j)}$. 
Here, the $\bar{\mathbf{x}}_{n-1,k}^{(j)}$ are obtained by sampling 
with replacement from the set $\big\{\mathbf{x}_{n-1,k}^{(j')}\big\}_{j'=1}^{J}$, 
where $\mathbf{x}_{n-1,k}^{(j')}$ is sampled with probability 
$w_{n-1,k}^{(j')}$. 

\vspace{1.7mm}
 
\item For each $\bar{\mathbf{x}}_{n-1,k}^{(j)}$, a new, ``predicted'' particle $\mathbf{x}_{n,k}^{(j)}\rmv$ is 
sampled from 
$f(\mathbf{x}_{n}|\mathbf{x}_{n-1})\big|_{\mathbf{x}_{n-1} =\, \bar{\mathbf{x}}_{n-1,k}^{(j)}}\!$. 

\vspace{1.2mm}

\item An approximation $\tilde{f}(\mathbf{z}_{n}|\mathbf{x}_{n})$ of the JLF $f(\mathbf{z}_{n}|\mathbf{x}_{n})$ is computed by means of LC
as described in Section \ref{sec:approx_calc_JLF}. This step requires 
communications with neighboring sensors. The local approximation at sensor $k$ can be calculated by means of LS fitting as described in Section \ref{sec:LS_Approx}, 
using the predicted particles $\big\{ \mathbf{x}_{n,k}^{(j)} \big\}_{j=1}^{J}$. 

\vspace{1mm}

\item The 
weights associated with the predicted particles $\mathbf{x}_{n,k}^{(j)}$ obtained in Step 2 are calculated according 
\vspace{-.5mm}
to 
\be
w_{n,k}^{(j)} \rmv\eq \frac{\tilde{f}(\mathbf{z}_{n}|\mathbf{x}_{n,k}^{(j)})}{\sum_{j'=1}^{J}\tilde{f}(\mathbf{z}_{n}|\mathbf{x}_{n,k}^{(j')})} \,, \quad\; j=1,\dots,J \,.
\label{eq:PF_weights}
\vspace{-.7mm}
\ee 
This involves the approximate JLF $\tilde{f}(\mathbf{z}_{n}|\mathbf{x}_{n})$ calculated in Step 3, which is evaluated at all predicted particles $\mathbf{x}_{n,k}^{(j)}$.

\vspace{1mm}

\item From 
$\big\{ \big( \mathbf{x}_{n,k}^{(j)}, w_{n,k}^{(j)} \big) \big\}_{j=1}^{J}$, an approximation of the global MMSE state estimate \eqref{eq:mmse_est} 
is computed according to \eqref{eq:pfEstimate_CPF}, 
\vspace{-1.5mm}
i.e.,
\[
\hat{\mathbf{x}}_{n,k} \rmv\eq\rmv \sum_{j=1}^{J} w_{n,k}^{(j)} \ist \mathbf{x}_{n,k}^{(j)} \,. 
\vspace{-.5mm}
\]
\end{enumerate}

The recursion defined by Steps 1--5 is initialized at 
$n\!=\!0$ by $J$ particles $\mathbf{x}_{0,k}^{(j)}$ 
sampled 
(at each sensor) from a suitable prior pdf $f(\mathbf{x}_{0})$, and by equal weights $w_{0,k}^{(j)} \equiv 1/J$. 
}
\vspace{1.7mm}
{\hrule}
\vspace{4mm}

Through the above recursion, each sensor obtains a global quasi-MMSE state estimate that involves the past and current measurements of all sensors.
Because of the use of LC, this is achieved without communicating 
between distant sensors or employing complex routing protocols. 
Also, no particles, local state estimates, or measurements need to be communicated between sensors. 
The local PF algorithms running at different sensors are identical. Therefore,    
any differences between the state estimates $\hat{\mathbf{x}}_{n,k}$ at different sensors $k$ are only due to the 
random sampling of the particles (using nonsynchronized local random generators) and errors caused by insufficiently 
converged consensus algorithms. 

\vspace{-1.3mm}

\subsection{Communication Requirements} \label{subsec:perf}

\vspace{.8mm}

We 
now discuss the 
communication requirements of our LC-based distributed PF (LC-DPF). 
For comparison, we also consider the centralized PF (CPF) of Section \ref{subsec:CPF}, in which all sensor measurements are transmitted to a fusion center (FC), 
and a straightforward distributed PF implementation (S-DPF) in which the measurements of each sensor are transmitted to all other sensors.
Note that with the S-DPF, each sensor performs exactly the same PF operations as does the FC in the CPF scheme. 

For 
the CPF, 
communicating all sensor measurements at time $n$ to the FC requires the transmission of
a total of $\sum_{k=1}^{K} H_{k} N_{n,k}$ real numbers within the sensor network \cite{coates2004distributed}. 
Here, $H_k$ denotes the number of communication hops from sensor $k$ to the FC, and $N_{n,k}$ is the dimension of $\mathbf{z}_{n,k}$. 
Additional information needs to be transmitted to the FC if the FC does not possess prior knowledge of the JLF. 
Finally, if the state estimate calculated at the FC is required to be available at the sensors, 
additional $MH^\prime$ 
real numbers need to be 
transmitted at each time $n$. Here, $H^\prime$ denotes the number of communication hops needed to disseminate the state estimate throughout the network. 
A problem of the CPF using multihop transmission 
is that all data pass through a small 
subset of sensors surrounding the FC, which can lead to fast depletion of the batteries of these sensors. 

With the S-DPF, disseminating the measurements of all sensors at time $n$ to all other sensors requires the transmission of $\sum_{k=1}^{K} H_{k}^{\prime\prime} N_{n,k}$ 
real numbers \cite{coates2004distributed}, where $H_{k}^{\prime\prime}$ is the number of communication hops required to disseminate the measurement of sensor $k$ throughout the network. 
Again, additional information needs to be transmitted if the JLF is not 
known to all sensors.

Finally, the proposed LC-DPF requires the transmission of $K I N_{c}$ real numbers at each time $n$, where 
$I$ is the number of consensus iterations performed by each consensus algorithm and $N_{c} = R_a + R_d$ is the number of consensus algorithms 
executed in parallel (see Section \ref{sec:approx_calc_JLF}). 
In contrast to the CPF and S-DPF, this number of transmissions 
does not depend on the measurement dimensions $N_{n,k}$; this 
makes the LC-DPF particularly attractive in the case of high-dimensional 
measurements. 
Another advantage of the LC-DPF is that no additional communications are needed (e.g., to transmit local likelihood functions between sensors).
Furthermore, the LC-DPF does not require multihop transmissions or routing protocols since each sensor simply broadcasts information to its neighbors.
This makes the LC-DPF particularly suited to wireless sensor networks with 
dynamic network topologies (e.g., moving sensors or a time-varying number of active sensors): in contrast to the CPF and S-DPF, there is no need to rebuild routing tables 
each time the network topology changes. 

On the other hand, the computational complexity of the LC-DPF is higher than that of the S-DPF because
the approximation described in Section \ref{sec:approx-exp} needs to be computed at each sensor. Overall, the LC-DPF performs more local computations than the S-DPF 
in order to reduce communications; this is especially true for 
high-dimensional measurements and/or high-dimensional parametrizations of the local likelihood functions.
Since the energy consumption 
of local computations is typically much smaller 
than that 
of communication, the total energy consumption is reduced and thus the operation lifetime is extended. This advantage of the LC-DPF comes 
at the cost 
of a certain performance loss (compared to the CPF or S-DPF) due to the approximate JLF used by 
the local PFs.
This will be analyzed experimentally in Section \ref{sec:sim-results}. 

\vspace{-.5mm}

\section{Distributed Gaussian Particle Filtering}\label{sec:distributed-gaussian-particle-filtering}

\vspace{.5mm}

Next, we propose two distributed versions of the \emph{Gaussian PF} (GPF). The GPF was introduced in \cite{kotecha2003gaussian}
as a simplified version of the PF using a Gaussian approximation of the posterior $f(\mathbf{x}_n|\mathbf{z}_{1:n})$. 
The mean and covariance of this Gaussian approximation are 
derived from a weighted particle set. The particles and their weights are computed in a similar way as 
described in Section \ref{sec:distributed-particle-filtering}, with the difference that no resampling is required. This results in a reduced complexity 
and allows for a parallel implementation \cite{bolic2009gpf}. 

\vspace{-1.5mm}

\subsection{Distributed Gaussian Particle Filtering Using LC}\label{sec:LGPF}

\vspace{.8mm}

In the proposed distributed GPF schemes, sensor $k$ uses a local GPF to 
track the mean vector $\bm{\mu}_{n,k}$ 
and covariance matrix $\mathbf{C}_{n,k}$ of a local Gaussian approximation $\mathcal{N}(\bm{\mu}_{n,k},\mathbf{C}_{n,k})$ 
of 
the global posterior $f(\mathbf{x}_n|\mathbf{z}_{1:n})$. 
The state estimate $\hat{\mathbf{x}}_{n,k}$ of sensor $k$ at time $n$ is defined to be the current mean, i.e, $\hat{\mathbf{x}}_{n,k} \!=\! \bm{\mu}_{n,k}$. 
The calculation of this estimate is based on the past and current measurements of all sensors, $\mathbf{z}_{1:n}$.
As with the distributed PF described in Section \ref{subsec:distributed-pf} (Algorithm 2), 
these measurements are epitomized by an approximation of the
JLF, which is provided to each sensor by means of LC. A statement of the algorithm follows.

\vspace{4mm}
{
{\hrule
\begin{center}\vspace{-1.2mm}
{\small\sc Algorithm 3:\, LC-based Distributed GPF (LC-DGPF)}\vspace{-1.2mm}
\end{center}
\hrule}
\vspace{2.8mm}

\small
\renewcommand{\baselinestretch}{1.18}\normalsize\small

\noindent
At time $n$, the local GPF at sensor $k$ performs the following steps, which are identical for all $k$. 

\vspace{1mm}

\begin{enumerate}

\item[1)] $J$ particles $\big\{ \bar{\mathbf{x}}_{n-1,k}^{(j)} \big\}_{j=1}^{J}$ are 
sampled from the previous local Gaussian approximation $\ist\mathcal{N}(\bm{\mu}_{n-1,k},\mathbf{C}_{n-1,k})$, 
where $\bm{\mu}_{n-1,k}$ and $\mathbf{C}_{n-1,k}$ were calculated at time $n\!-\!1$. Note that this sampling step 
replaces the resampling step of the distributed PF of Section \ref{subsec:distributed-pf} (Step 1 in 
Algorithm 2).

\vspace{1.3mm}

\item[2)--4)] These steps 
are identical to the corresponding steps of the distributed PF of 
Section \ref{subsec:distributed-pf} (Algorithm 2); they
involve LC (Step 3) and result in a set of ``predicted'' particles and corresponding weights, $\big\{ \big( \mathbf{x}_{n,k}^{(j)}, w_{n,k}^{(j)} \big) \big\}_{j=1}^{J}$.

\vspace{1mm}

\item[5)] From $\big\{ \big( \mathbf{x}_{n,k}^{(j)}, w_{n,k}^{(j)} \big) \big\}_{j=1}^{J}$, the mean $\bm{\mu}_{n,k}$ and 
covariance $\mathbf{C}_{n,k}$ of the Gaussian approximation $\mathcal{N}(\bm{\mu}_{n,k},\mathbf{C}_{n,k})$
of 
the current posterior $f(\mathbf{x}_{n}|\mathbf{z}_{1:n})$ are calculated 
\vspace{-.5mm}
as
\be
\begin{array}{l}
\displaystyle {\bm{\mu}}_{n,k} \eq \sum_{j=1}^{J} w_{n,k}^{(j)} \ist \mathbf{x}_{n,k}^{(j)}\\[5mm]
\displaystyle {\mathbf{C}}_{n,k} \eq \sum_{j=1}^{J} w_{n,k}^{(j)} \ist \mathbf{x}_{n,k}^{(j)}\mathbf{x}_{n,k}^{(j)\top} \rmv-\ist \bm{\mu}_{n,k} \ist \bm{\mu}_{n,k}^{\top} \,.
\end{array} 
\label{eq:muAndC}
\vspace{-.5mm}
\ee
The 
state estimate $\hat{\mathbf{x}}_{n,k}$ (approximating $\hat{\mathbf{x}}_{n}^{\text{MMSE}}$ in \eqref{eq:mmse_est}) is taken to be 
the posterior 
\vspace{.5mm}
mean $\bm{\mu}_{n,k}$. 

\end{enumerate}

The recursion defined by Steps 1--5 is initialized as in Algorithm 2. 
}
\vspace{2mm}
{\hrule}

\subsection{Reduced-Complexity Method} \label{sec:LC-DGPF_red}  

\vspace{.8mm}

We next present a reduced-complexity variant of the LC-DGPF described above, 
in which each of the $K$ local GPFs uses only $J' \rmv \triangleq J/K$ particles. Here, $J$ is chosen such that $J'$ is an integer and 
$J ' \!\ge \max \{R_a,R_d\}$ (cf.\ Section \ref{sec:LS_Approx}). 
The sets of $J'$ particles of all local GPFs are effectively combined via a second stage of consensus algorithms, 
such that a ``virtual global GPF'' with $J \!=\! KJ'\rmv$ particles is obtained. In other words, $J$ particles---which, in the LC-DGPF, 
were used by each individual sensor separately---are ``distributed'' over the $K$ sensors. 
As a consequence, the computational complexity of the local GPFs is substantially reduced while the estimation accuracy 
remains effectively unchanged.
This advantage comes at the cost of some increase in local communications due to the additional 
consensus algorithms. 

This reduced-complexity method is similar to a parallel GPF implementation 
proposed in \cite{bolic2009gpf}, which uses 
multiple processing units---corresponding to our sensors---collocated with a central unit. However, instead of a central unit, 
we 
employ distributed consensus algorithms to combine the partial estimates (means) and partial covariances 
calculated at the individual sensors. Another difference from \cite{bolic2009gpf} is the use of 
an approximate JLF that is obtained in a distributed way by means of LC. 
The algorithm is stated 
\vspace{.5mm}
as follows.

\vspace{5mm}
{
{\hrule
\begin{center}\vspace{-.8mm}
{\small\sc Algorithm 4:\, Reduced-complexity LC-DGPF\\[-.5mm]
(R-LC-DGPF)}\vspace{-1mm}
\end{center}
\hrule}
\vspace{2.8mm}

\small
\renewcommand{\baselinestretch}{1.18}\normalsize\small

\noindent
At time $n$, the local GPF at sensor $k$ first performs Steps 1--3 of the LC-DGPF algorithm described in Section \ref{sec:LGPF} (Algorithm 3), 
however using $J' \rmv = J/K$ rather than $J$ particles. 
The remaining steps, described in the following, 
are modified versions of Steps 4 and 5 of 
Algorithm 3, as well as an additional consensus step. 

\vspace{1mm}

\begin{enumerate}

\item[4)] Nonnormalized weights are calculated as 
(cf.\ \eqref{eq:PF_weights})
\[
\tilde{w}_{n,k}^{(j)} \rmv\eq\rmv \tilde{f}(\mathbf{z}_{n}|\mathbf{x}_{n,k}^{(j)}) \,, \quad\; j=1,\dots,J' .
\] 
This requires evaluation of the approximate JLF $\tilde{f}(\mathbf{z}_{n}|\mathbf{x}_{n})$, which was calculated in Step 3 using
LC, at the $J'$ predicted particles $\big\{ \mathbf{x}_{n,k}^{(j)} \big\}_{j=1}^{J'}$ drawn in Step 2. 
Furthermore, the sum of the $J'\rmv$ nonnormalized weights is 
\vspace{-.5mm}
computed: 
\[
\widetilde{W}_{n,k} \rmv\eq\rmv \sum_{j=1}^{J'} \tilde{w}_{n,k}^{(j)} \,.
\vspace{-.3mm}
\] 

\item[5)] From the weighted particles $\big\{ \big( \mathbf{x}_{n,k}^{(j)}, \tilde{w}_{n,k}^{(j)} \big) \big\}_{j=1}^{J'}$, a partial nonnormalized mean 
and a partial nonnormalized correlation
are calculated 
as
\be
{\bm{\mu}}'_{n,k} \rmv\eq\rmv \sum_{j=1}^{J'} \tilde{w}_{n,k}^{(j)} \ist \mathbf{x}_{n,k}^{(j)} \,, \quad\;
{\mathbf{R}}'_{n,k} \rmv\eq\rmv \sum_{j=1}^{J'} \tilde{w}_{n,k}^{(j)} \ist \mathbf{x}_{n,k}^{(j)}\mathbf{x}_{n,k}^{(j)\top}, 
\label{eq:partial_m_C}
\vspace{-.3mm}
\ee
respectively. Note that Steps 4 and 5 are carried out locally at sensor $k$.

\vspace{1.5mm}

\item[6)] The partial means and correlations from all sensors are combined to obtain the global mean and covariance: 
\be
{\bm{\mu}}_{n} \rmv\eq\rmv \frac{1}{W_n} \sum_{k=1}^{K} \bm{\mu}'_{n,k}\,, \quad\;
{\mathbf{C}}_{n} \rmv\eq\rmv \frac{1}{W_n} \sum_{k=1}^{K} {\mathbf{R}}'_{n,k} \rmv- \bm{\mu}_{n}\bm{\mu}_{n}^{\top} \ist,
\label{eq:muAndC_complete}
\ee
where 
\vspace{-.5mm}
\begin{equation}
W_n \rmv\eq\rmv \sum_{k=1}^{K}\widetilde{W}_{n,k}
\label{eq:W_complete} 
\end{equation}
is the global sum of all particle weights. The sums over all sensors in \eqref{eq:muAndC_complete} and \eqref{eq:W_complete} 
are computed in a distributed manner by means of consensus algorithms. 
The normalization by $W_n$ and subtraction of $\bm{\mu}_{n}\bm{\mu}_{n}^{\top}$ in \eqref{eq:muAndC_complete} are performed locally 
at each sensor after convergence of these consensus algorithms.
The state estimate $\hat{\mathbf{x}}_{n}$ 
is taken to be $\bm{\mu}_{n}$.
\end{enumerate}
}
\vspace{2mm}
{\hrule}
\vspace{4mm}

As a result of this algorithm, all sensors obtain identical $\hat{\mathbf{x}}_{n} = \bm{\mu}_{n}$ and $\mathbf{C}_{n}$ 
provided that the consensus algorithms are sufficiently converged.
Therefore, 
we omit the subscript $k$ indicating the sensor dependence (cf. \eqref{eq:muAndC}), 
i.e., we write $\hat{\mathbf{x}}_{n} = \bm{\mu}_{n}$ instead of $\hat{\mathbf{x}}_{n,k} = \bm{\mu}_{n,k}$ and $\mathbf{C}_{n}$ instead of 
$\mathbf{C}_{n,k}$ for all $k$.

It is easily seen from 
\eqref{eq:partial_m_C}--\eqref{eq:W_complete} that $\bm{\mu}_{n}$ and $\mathbf{C}_{n}$ are actually the result of an averaging (summation) over 
$J$ particles (note that $J' \rmv= J/K$ particles are sampled independently at each of the $K$ sensors). 
Therefore, under the assumption that the consensus algorithms used to calculate the sums over all sensors in \eqref{eq:muAndC_complete} 
and \eqref{eq:W_complete} are converged, 
$\bm{\mu}_{n}$ and $\mathbf{C}_{n}$ should ideally be effectively equal to the corresponding quantities obtained by the LC-DGPF. 
However, a certain performance degradation is caused by the fact that the LS fitting performed at each sensor (see Section \ref{sec:LS_Approx}) 
is now based on only $J' \rmv = J/K$ predicted particles $\mathbf{x}_{n,k}^{(j)}$, and hence the resulting approximate local likelihood functions and, in turn, 
the approximate JLF will be less accurate. In Section \ref{sec:sim-results}, we will show by means of simulations that this degradation is very small.

\subsection{Computational Complexity and Communication Requirements} \label{sec:perf_LC-DGPF}

\vspace{.8mm}

We compare the computational complexity and communication requirements of the LC-DGPF
and of its reduced-complexity variant discussed above (abbreviated R-LC-DGPF). 
We will disregard Steps 2 and 3 of the LC component 
(Algorithm 1), because  
their complexity and communication requirements 
are identical for the LC-DGPF and R-LC-DGPF; furthermore, their 
complexity 
is 
typically\footnote{The 
complexity of Steps 2 and 3 of Algorithm 1 is linear in the number of consensus algorithms and in the number of consensus iterations; these numbers depend
on the specific application and 
setting.} 
much lower than that 
of the remaining steps 
(local GPF algorithm and LS approximation). 

The complexity of the local GPF algorithm and of the LS approximation in the LC scheme (Step 1 of Algorithm 1) depends linearly on the 
number of particles \cite{bolic2009gpf, bjorck1996numerical}. Thus, reducing the number of particles at each sensor from $J$ to $J' \!=\rmv J/K$ 
reduces this complexity by a factor of $K$. It follows that the R-LC-DGPF is significantly less complex than the LC-DGPF. 
(The complexity 
of the additional consensus algorithms required by the R-LC-DGPF 
is typically negligible compared to the other operations.) 
The additional communication requirements of the R-LC-DGPF relative to the LC-DGPF
are determined primarily by the speed of convergence (i.e., number of iterations $I$) of the additional consensus algorithms, 
which depends mainly on the second smallest eigenvalue of the Laplacian of the communication graph \cite{fiedler1973algebraic}, 
and by 
the state dimension $M$. More specifically, 
the additional number of real numbers transmitted in the entire sensor network at each time $n$ 
is $K I N_{c}'$, where 
$N_c' = M + M(M \rmv+\rmv 1)/2 + 1$ is 
the number of additional consensus algorithms, i.e., 
of (scalar) consensus algorithms needed to calculate the mean vector
and covariance matrix in \eqref{eq:muAndC_complete} as well as the total weight in \eqref{eq:W_complete}. 
{Since $N_c'$ is of order $M^2$\!, the R-LC-DGPF has a disadvantage 
for high-dimensional states.} 

The reduced operation count of the R-LC-DGPF relative to the LC-DGPF can be exploited in two alternative ways, which represent 
a tradeoff between latency and power consumption. First, 
the processing time can be reduced; this 
results in a smaller 
latency of the R-LC-DGPF relative to the LC-DGPF, provided that the delays caused by the additional consensus algorithms are not too large.
Thus, the R-LC-DGPF may be more suitable for real-time applications; however, the power consumption is higher due to the increased 
communications. 
Alternatively, if latency is not an issue, the processor's clock frequency can be reduced. 
The processing time can then be made equal to that of the LC-DGPF, 
while 
the processor's power consumption is reduced due to the lower clock frequency \cite{wang2002energy}.
Thereby, the overall power consumption of the R-LC-DGPF is smaller 
relative to the LC-DGPF, provided that the additional power consumption due to the increased communications 
is not too large. However, the total latency is increased by the delays caused by the additional consensus algorithms. 

\vspace{-4mm}

\section{Numerical Study}\label{sec:sim-results}

\vspace{.5mm}

We will now apply the proposed LC-based distributed PF algorithms to the problem of tracking multiple targets using acoustic amplitude 
sensors. 
We will compare the performance of our methods with that of the centralized PF 
and state-of-the-art distributed PFs. 

\vspace{-1mm}

\subsection{Acoustic-Amplitude-Based Multiple Target Tracking}\label{subsec:target-tracking}  

\vspace{.8mm}

We consider $P$ targets 
($P$ assumed known) moving independently in the $x$-$y$ plane. 
The $p\ist$th target, $p \in \{1,\ldots,P\}$, 
is represented by the state vector  $\mathbf{x}_{n}^{(p)} \!\triangleq \big(x_{n}^{(p)} \;\ist y_{n}^{(p)} \;\ist \dot{x}_{n}^{(p)} \;\ist \dot{y}_{n}^{(p)} \big)^{\top}\!$ 
containing the target's 2D position and 2D velocity. The overall state vector is defined as 
$\mathbf{x}_{n} \triangleq \big(\mathbf{x}_{n}^{(1)\top}\!\! \cdots\,\mathbf{x}_{n}^{(P)\top}\big)^{\top}\!$. 
Each vector 
$\mathbf{x}_{n}^{(p)}$ evolves 
independently of the other vectors $\mathbf{x}_{n}^{(p')}$ according to 
$\mathbf{x}_{n}^{(p)} \!= \mathbf{G}_{p}\mathbf{x}_{n-1}^{(p)} + \mathbf{W}_{p}\mathbf{u}_{n}^{(p)}$. Here,
$\mathbf{u}_{n}^{(p)} \rmv\sim\rmv \mathcal{N}(\mathbf{0}_2,\sigma_u^2 \ist \mathbf{I}_2)$ is Gaussian driving noise, 
with $\mathbf{u}_{n}^{(p)}$ and $\mathbf{u}_{n'}^{(p')}$ independent unless $(n,p) \!=\! (n'\rmv,p')$, 
and $\mathbf{G}_{p}\!\in\! \mathbb{R}^{4\times 4}$ and $\mathbf{W}_{p}\!\in\! \mathbb{R}^{4\times 2}$ are system matrices that 
will be specified in Section \ref{subsec:simu-settings}. This model is commonly used in target tracking applications 
\cite{kotecha2003gaussian,djuric2008target, bar2001estimation, gustafsson2002particle}. 
It follows that the overall 
state vector $\mathbf{x}_{n}$ evolves 
according to 
\[
\mathbf{x}_{n} =\, \mathbf{G}\mathbf{x}_{n-1} + \mathbf{W}\mathbf{u}_{n} \,, \quad\; n=1,2,\dots \,, 
\]
where $\mathbf{G} \!\triangleq\rmv \mathrm{diag} \ist\{\mathbf{G}_{1},\dots,\mathbf{G}_{P}\}$, 
$\mathbf{W} \!\triangleq\rmv \mathrm{diag} \ist\{\mathbf{W}_{1},\dots,\mathbf{W}_{P}\}$, 
and $\mathbf{u}_{n} \triangleq \big(\mathbf{u}_{n}^{(1)\top}\!\!\cdots \mathbf{u}_{n}^{(P)\top}\big)^{\top} 
\!\!\sim\rmv \mathcal{N}(\mathbf{0}_{2P},\sigma_u^2 \ist \mathbf{I}_{2P})$. 

Each target $p$ emits a sound with a (root mean-squared) amplitude 
$A_p$ that is assumed constant and known. 
At the position of sensor $k$, denoted $\bm{\xi}_{n,k}\rmv$, the sound amplitude due to target $p$ 
is modeled as $A_p/\|\bm{\rho}_{n}^{(p)} \rmv-{\bm{\xi}_{n,k} \|}^{\kappa}$, where $\bm{\rho}_{n}^{(p)} \!\triangleq \big(x_{n}^{(p)} \;\ist y_{n}^{(p)} \big)^{\top}\rmv$ is the position of 
target $p$ and $\kappa$ is the path loss exponent \cite{djuric2008target ,sheng2004maximum,liu2003cnp}. 
The (scalar) measurement $z_{n,k}\rmv$ obtained by sensor $k$ at time $n$ is then given by 
\begin{align}
&z_{n,k} \eq h_{n,k}(\mathbf{x}_n) \ist +\ist v_{n,k} \,, \nonumber\\[.5mm]
&\rule{15mm}{0mm}\text{with} \;\; 
h_{n,k}(\mathbf{x}_n) \eq \sum_{p=1}^{P}\frac{A_p}{ \|\bm{\rho}_{n}^{(p)} \rmv-\bm{\xi}_{n,k} \|^{\kappa} } \,, 
\label{eq:measModel_tracking}
\end{align}
where $v_{n,k} \rmv\sim\mathcal{N}(0,\sigma_{v}^{2})$ are zero-mean Gaussian measurement noise variables of equal variance $\sigma_{v}^{2}$. We assume that $v_{n,k}$ is independent of $\mathbf{x}_{n'}$ for all $n'$, and that $v_{n,k}$ and $v_{n'\!,k'}$ are independent unless $(n,k) = (n'\!,k')$. 
Note that this measurement model is a special instance of \eqref{eq:scalar_meas_model_additive_gauss},
and that $z_{n,k}$ does not depend on the velocities $\dot{x}_{n}^{(p)}$ and $\dot{y}_{n}^{(p)}$.
The local likelihood functions and the JLF are respectively 
given by 
(cf.\ \eqref{eq:local_likelihood_additive_gauss}, \eqref{eq:JLF_additive_gauss}) 
\begin{align}
\hspace*{-2.5mm}f(z_{n,k}|\mathbf{x}_n) &\eq \frac{1}{\sqrt{2\pi \sigma_{v}^{2}}} \, \exp\!\bigg(\!\!-\rmv \frac{1}{2\sigma_{v}^{2}} \ist [z_{n,k} \!-\rmv h_{n,k}(\mathbf{x}_n)]^2 \rmv \bigg)   \label{eq:LLF_gauss}\\[0mm]
\hspace*{-2.5mm}f(\mathbf{z}_{n}|\mathbf{x}_n) &\eq \frac{1}{\sqrt{(2\pi \sigma_{v}^{2})^K}} \, \exp\!\Bigg(\!\!-\rmv \frac{1}{2\sigma_{v}^{2}} \ist 
  \sum_{k=1}^K [z_{n,k} \!-\rmv h_{n,k}(\mathbf{x}_n)]^2 \rmv \Bigg) \,,
  \nonumber
\end{align}
and hence (cf.\ \eqref{eq:sum_Sz_exp_gauss})
\[
S_n(\mathbf{z}_n,\mathbf{x}_n) \eq \frac{1}{\sigma_{v}^{2}} \sum_{k=1}^{K} h_{n,k}(\mathbf{x}_n) \ist \bigg[ z_{n,k} - \frac{1}{2} \ist h_{n,k}(\mathbf{x}_n) \bigg] \,,
\]
with $h_{n,k}(\mathbf{x}_n)$ given by \eqref{eq:measModel_tracking}.

In general, the sensor positions $\bm{\xi}_{n,k}$ are allowed to change with time $n$.
(However, 
we used static sensors for simplicity.)
Each sensor is supposed to know its own position but 
not the positions of the other sensors.
The sensor positions (which are contained in the local likelihood functions) are implicitly fused by the LC method in the process of calculating the JLF; 
they need not be explicitly transmitted between the sensors. 
Therefore, the LC method and 
our LC-based distributed (G)PFs 
are well suited for dynamic sensor networks. 

\begin{figure*}[t]
\centering


\psfrag{true}[][][0.62]{\hspace{18.5mm}\raisebox{0mm}{True trajectories}}
\psfrag{tracked}[][][0.62]{\hspace{17.1mm}\raisebox{0mm}{Tracked trajectories}}

\psfrag{0t}[][][0.63]{\hspace{15mm}\raisebox{25mm}{{\huge$\bm{\star}$}}}
\psfrag{5t}[][][0.63]{\hspace{126mm}\raisebox{145mm}{{\huge$\bm{\star}$}}}
\psfrag{50s}[][][0.63]{\hspace{68mm}\raisebox{110mm}{{\textcolor{blue}{\large$\bm{\blacksquare}$}}}}

\psfrag{n}[][][0.80]{\hspace{-3mm}\raisebox{-7mm}{$n$}}
\psfrag{RMSE}[][][0.80]{\hspace{0mm}\raisebox{8mm}{$\text{RMSE}_n$  [m]}}

\hspace*{-.5mm}
\subfigure[ ]{
\psfrag{5}[][][0.63]{\hspace{15mm}\raisebox{25mm}{{\huge$\bm{\star}$}}}
\includegraphics[height=5cm,width=5cm]{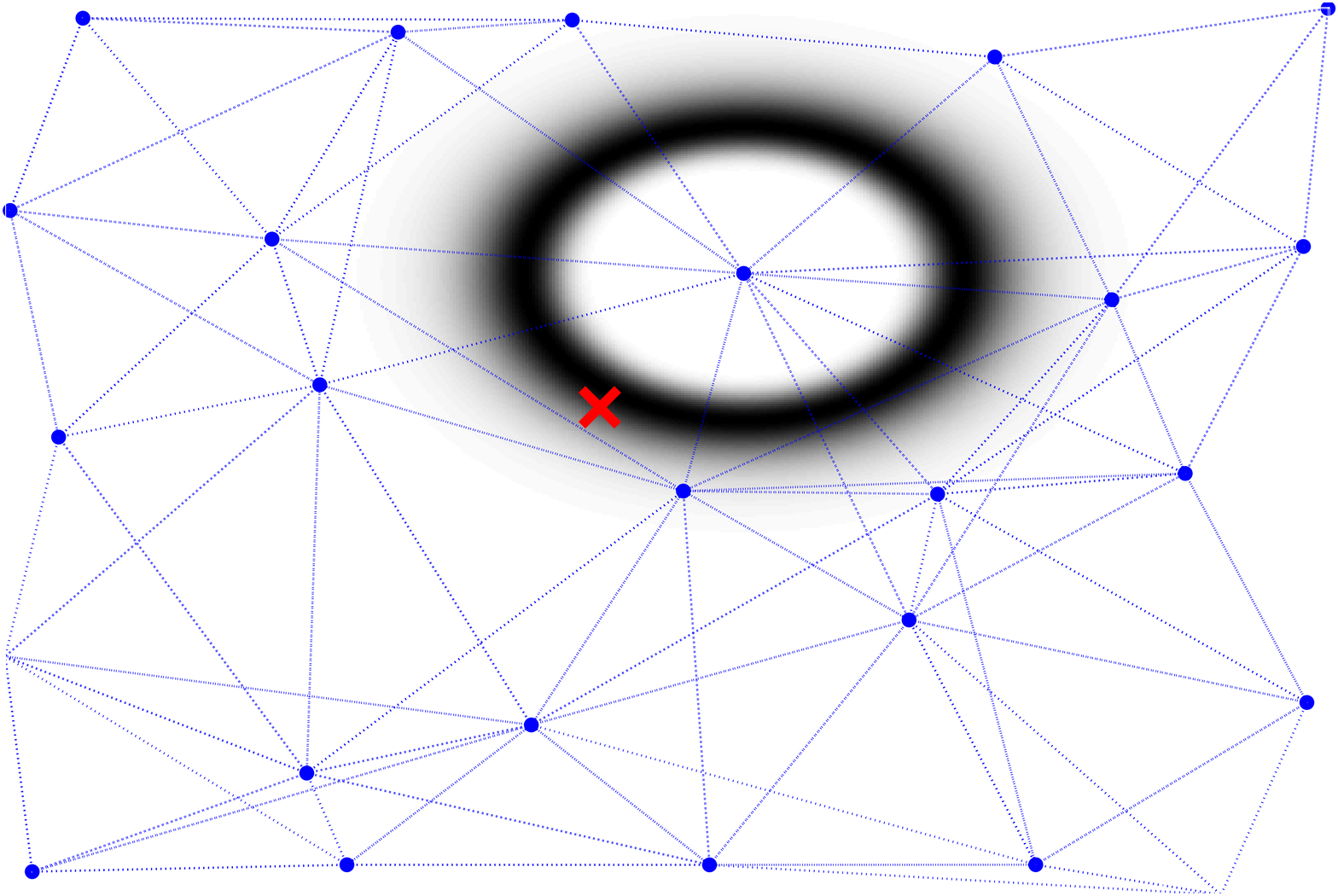}
}
\hspace{5mm}
\subfigure[ ]{
\includegraphics[height=5cm,width=5cm]{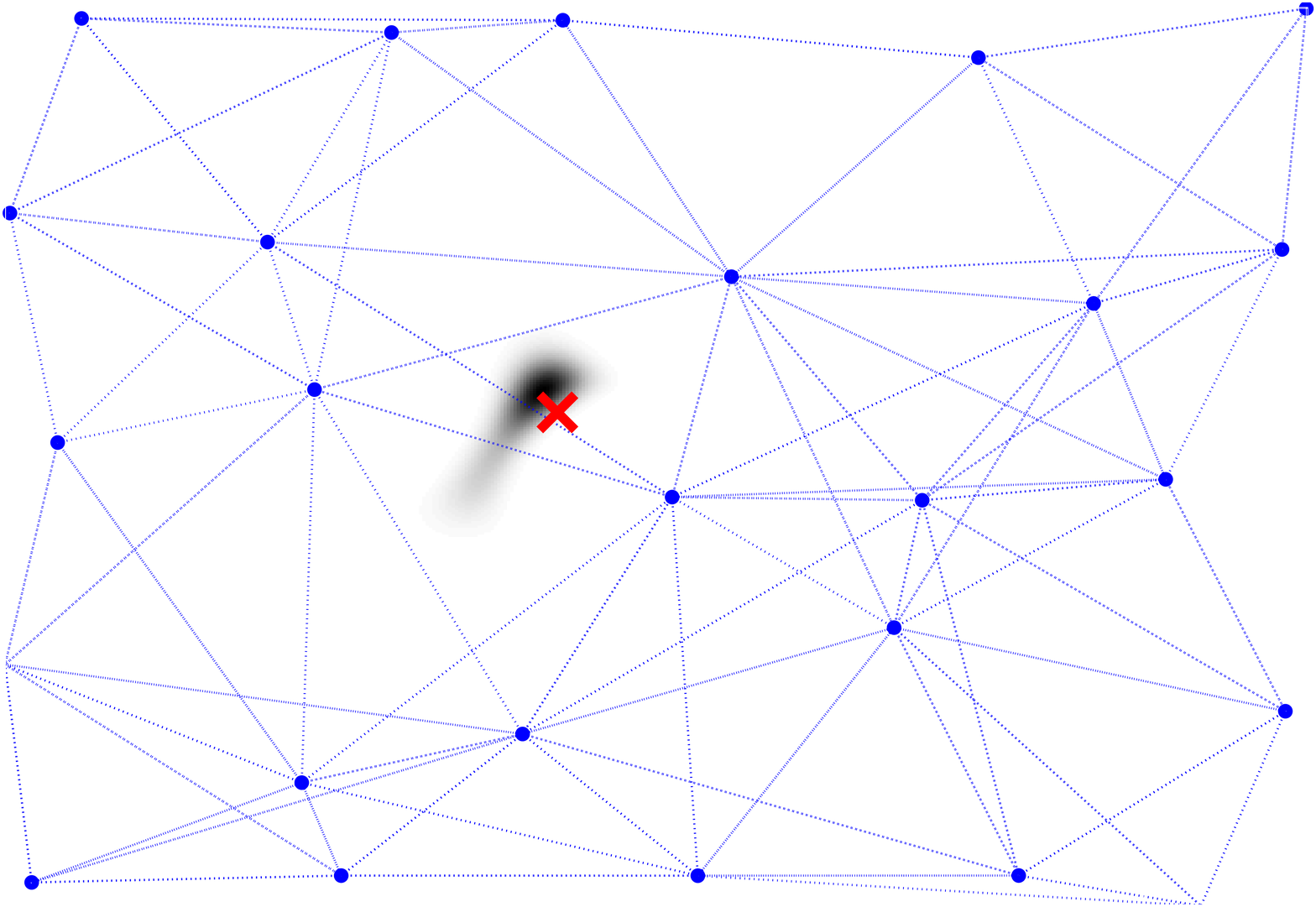}
}
\hspace{5mm}
\subfigure[ ]{
\includegraphics[height=5cm,width=5cm]{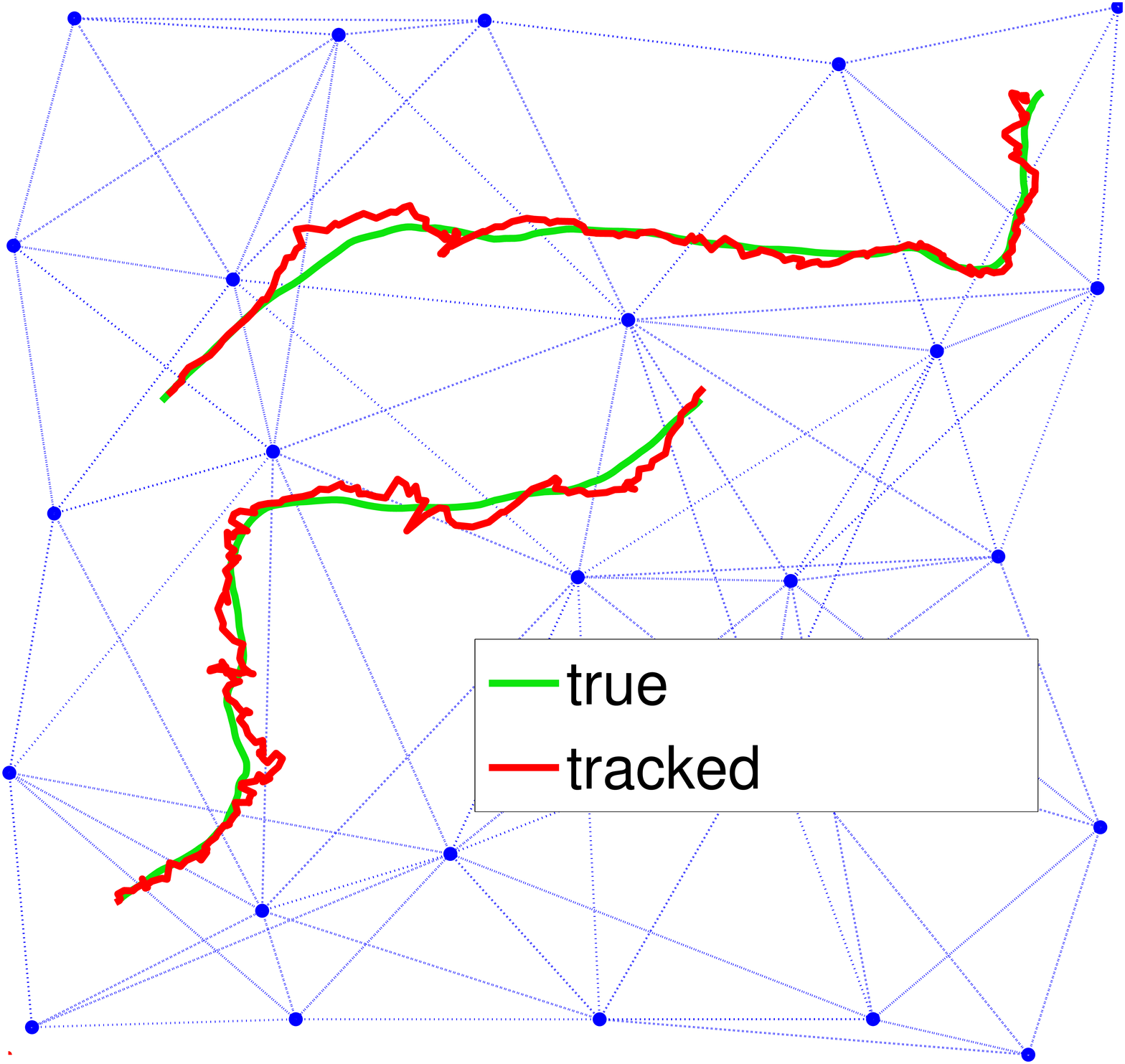}
}
\vspace*{0mm}
\renewcommand{\baselinestretch}{1.2}\small\normalsize
\caption{Example of a sensor network and communication topology, along with 
{(a) a local likelihood function for one target, 
(b) a JLF for one target,} and 
(c) a realization of the trajectories of two targets and the corresponding trajectories
tracked by the LC-DPF. 
{In (a), the square indicates the sensor for which the local likelihood is depicted. 
In (a) and (b), darker shading represents higher likelihood values and the cross indicates 
the position of the target.
In (c), the stars indicate the start points of the target trajectories.}} 
\label{fig:trajectories}
\vspace{0mm}
\end{figure*}

\vspace{-.8mm}

\subsection{Simulation Setting} \label{subsec:simu-settings}

\vspace{.8mm}

In our 
simulations, the number of targets is $P \!=\! 2$ unless stated otherwise. The system matrices $\mathbf{G}_p$ and $\mathbf{W}_p$ 
are identical for the two targets and 
given by \cite{kotecha2003gaussian} 
\[
\mathbf{G}_p = \left(\!\! \begin{array}{cccc}
1 \!&\! 0 \!&\! 1 \!&\! 0 \\[-0mm]
0 \!&\! 1 \!&\! 0 \!&\! 1 \\[-0mm]
0 \!&\! 0 \!&\! 1 \!&\! 0 \\[-0mm]
0 \!&\! 0 \!&\! 0 \!&\! 1 \end{array}\!\!\!\right),
\quad
\mathbf{W}_p = \left(\!\! \begin{array}{cc}
0.5 \!&\! 0 \\[-0mm]
0 \!&\! 0.5 \\[-0mm]
1 \!&\! 0 \\[-0mm]
0 \!&\! 1 \end{array}\!\!\!\right) , \quad\; p = 1,2 \,.
\vspace{-.5mm}
\]
The variance of the driving noises $\mathbf{u}_{n}^{(p)}$ is given by $\sigma_{u}^{2} = 0.00035$. Each of the two targets emits a sound of equal amplitude 
$A_p \!=\! 10$. The initial prior pdf $f(\mathbf{x}_{0}^{(p)}) \!=\! \mathcal{N}(\bm{\mu}_0^{(p)},\mathbf{C}_0)$ is 
different for the two targets, with $\bm{\mu}_0^{(1)} = (36\; 36\; {-0.05}\; {-0.05})^{\top}\!$ for target 1, 
$\bm{\mu}_0^{(2)}=(4 \; 4\; 0.05 \; 0.05)^{\top}\rmv$ for target 2, and $\mathbf{C}_0 \!=\rmv \mathrm{diag} \ist\{1,1,0.001,0.001\}$ for both targets.

The network consists of $K \!\rmv=\! 25$ acoustic amplitude sensors that are deployed on a jittered grid 
within a rectangular region of size $40\ist\text{m} \times 40\ist\text{m}$. 
Each sensor communicates with other sensors within a range of $18\ist$m. The measurement noise variance is 
$\sigma^2_v \rmv=\rmv 0.05$ and the path loss exponent is $\kappa \rmv=\rmv 1$. 

\begin{table*}
\centering
\begin{tabular}{|c||c|c|c|c|c|c|}
  \hline
&  & \rule{0mm}{3.2mm}Track loss adjusted &  & Track loss adjusted & Track loss & Communication    \\[-.01mm]
& {ARMSE [m]} & ARMSE [m] & {$\sigma_{\text{ARMSE}}\,$[m]} & $\sigma_{\text{ARMSE}}\,$[m]  & percentage [\%] &  requirements\\[.8mm]
  \hline \hline
  \rule{0mm}{3mm}LC-DPF & {0.6225} & 0.5424 & {0.0860} & 0.0222 & 0.95 & 13800 \\[.2mm]
  \hline
  \rule{0mm}{3mm}LC-DGPF & {0.6187} & 0.5387 & {0.0889} & 0.0205 & 0.7 & 13800 \\[.2mm]
  \hline
  \rule{0mm}{3mm}R-LC-DGPF & {0.5531} & 0.5204 & {0.0005} & 0.0005 & 0.46 & 22800 \\[.2mm]
  \hline
  \rule{0mm}{3mm}GSHL-DPF \cite{gu2008consensus} & {1.3022} & 1.2841 & {0.0032} & 0.0032 & 0.74 & 8800 \\[.2mm]
  \hline
  \rule{0mm}{3mm}OC-DPF \cite{oreshkin2010async} & {0.9992} & 0.8399 & {0.0022} & 0.0024 & 1.1 & 8800 \\[.2mm]
  \hline
  \rule{0mm}{3mm}FRG-DPF \cite{farahmand2011set} & {0.5553} & 0.5335 & {0} & 0 & 0.2 & 400000 \\[.2mm]
  \hline     
  \rule{0mm}{3mm}CPF & {0.4975} & 0.4975 & {--} & -- & 0 & 770 \\[.2mm]
  \hline
  \rule{0mm}{3mm}CGPF & {0.5156} & 0.5086 & {--} & -- & 0.18 & 770 \\[.2mm]
  \hline
\end{tabular}
\vspace{2mm}
\renewcommand{\baselinestretch}{.9}\small\normalsize
\caption{Estimation performance and communication requirements of the proposed consensus-based distributed PFs (LC-DPF, LC-DGPF, 
and R-LC-DGPF), of state-of-the-art consensus-based distributed PFs (GSHL-DPF, OC-DPF, and FRG-DPF),
and of centralized PFs (CPF and CGPF).}
\vspace{-5mm}
\label{tab:compar}
\end{table*}

For LC, we approximate 
the measurement function 
$h_{n,k}(\mathbf{x}_n)$ in \eqref{eq:measModel_tracking} by a polynomial (see \eqref{eq:polynomial_approx}) of degree $R_p \!=\! 2$. 
This results in the following approximation of $S_n(\mathbf{z}_n,\mathbf{x}_n)$ 
(cf.\ \eqref{eq:approx_sum_exp_3_poly}):
\begin{align*}
&\tilde{S}_n(\mathbf{z}_n,\mathbf{x}_n)\nonumber\\[0mm]
&\rule{2mm}{0mm}\eq \sum_{k=1}^{25}\sum_{\mathbf{r}=\mathbf{0}}^{4} \beta_{n,k,\mathbf{r}}({z}_{n,k}) \,\ist (x_{n}^{(1)})^{r_1} \ist (y_{n}^{(1)})^{r_2}(x_{n}^{(2)})^{r_3} \ist (y_{n}^{(2)})^{r_4} \, .
\end{align*}
To obtain the  approximation coefficients ${\alpha}_{n,k,\mathbf{r}}$ needed for calculating the $\beta_{n,k,\mathbf{r}}({z}_{n,k})$
according to \eqref{eq:beta_exp} and \eqref{eq:gamma_exp}, we use LS fitting as described in Section \ref{sec:LS_Approx}. 
The sums over all sensors in \eqref{eq:statistic_poly} are computed by average consensus algorithms using 
Metropolis weights 
\cite{xiao2005scheme}.  
There are $N_c = {{\ist 4 \ist+\ist 4\ist} \choose {4}} - 1 = 69$ consensus algorithms that are executed in parallel, each using $I \!=\! 8$ 
iterations unless noted otherwise. 
The same remarks apply to the sums in \eqref{eq:muAndC_complete} and \eqref{eq:W_complete}, which are required by the R-LC-DGPF. 
The number of additional consensus algorithms employed by the R-LC-DGPF is $N_c' = 8 + 8 \rmv\cdot\rmv 9/2 + 1 = 45$.

We compare the LC-DPF, LC-DGPF, R-LC-DGPF, CPF, and a centralized GPF (CGPF), which, similarly to the CPF, processes all sensor measurements 
at an FC. In addition, we consider the state-of-the-art consensus-based distributed PFs proposed 
(i) by Gu et al.\ in \cite{gu2008consensus} (abbreviated GSHL-DPF), 
(ii) by Oreshkin and Coates in \cite{oreshkin2010async} (OC-DPF), and 
(iii) by Farahmand et al.\ in \cite{farahmand2011set} (FRG-DPF). 
Unless stated otherwise, the number of particles at each sensor was 
$J \!=\! 5000$ for the LC-DPF, LC-DGPF, GSHL-DPF, and OC-DPF; 
$J \!=\! 2000$ for the FRG-DPF (this reduction is made possible by the adapted proposal distribution); and
$J' \!=\! 5000/25 \rmv=\rmv 200$ for the R-LC-DGPF.
The PF at the FC of the CPF and CGPF employed $5000$ particles. 
In the FRG-DPF \cite{farahmand2011set}, the rejection probability used for proposal adaptation was set to $\beta_k=0.02$, 
and the oversampling factor was chosen as $L=10$. 

As a performance measure, we use the $n$-dependent 
root-mean-square error of the targets' position estimate $\hat{\bm{\rho}}_{n,k}$, denoted $\text{RMSE}_n$, which is computed as the square root of the 
average of 
$\big\| \hat{\bm{\rho}}_{n,k}^{(p)} \rmv-\rmv \bm{\rho}_n^{(p)} \big\|^2\rmv$ over the two targets $p=1,2$, all sensors $k = 1,\ldots,25$, and 
5000 simulation runs. 
Here, 
${\bm{\rho}}_{n}^{(p)}$ denotes the 
position of target $p$ and $\hat{\bm{\rho}}_{n,k}^{(p)}$ denotes the corresponding estimate 
at sensor $k$. 
We also compute the \emph{average} RMSE (ARMSE) 
as the square root of the average of $\text{RMSE}_n^2$ over all 200 simulated time instants $n$. 
Finally, we assess the error variation across the sensors $k$ by the standard deviation $\sigma_{\text{ARMSE}}$ 
of a $k$-dependent error 
defined as the square root of the average of $\big\| \hat{\bm{\rho}}_{n,k}^{(p)} \rmv-\rmv \bm{\rho}_n^{(p)} \big\|^2\rmv$ 
over the two targets $p=1,2$, all 200 time instants $n$, and 5000 simulation runs. 

\vspace{-.8mm}

\subsection{Simulation Results}\label{subsec:sim-results}

\vspace{.8mm}

Fig. \ref{fig:trajectories} shows an example of a sensor network and communication topology. 
{For the case of a single target ($P\!=\!1$), examples of the local likelihood function and of the JLF are visualized in Fig.\ \ref{fig:trajectories}(a) and (b), respectively. 
The local likelihood function is circularly symmetric because the measurement function $h_{n,k}(\mathbf{x}_n)$ in \eqref{eq:measModel_tracking} depends only on the distance between the target and the sensor. We can also see that the JLF is unimodal, which is an expected result since the JLF is the product of the 
local likelihood functions of all $K \!=\! 25$ sensors (see \eqref{eq:joint_likelihood_prod}), all having 
circularly symmetric shapes as shown in Fig.\ \ref{fig:trajectories}(a) but different locations due to  
the different local measurements and the different distances between target 
and sensor (see \eqref{eq:LLF_gauss}). Furthermore, we note that the nonlinearity of the local measurement functions $h_{n,k}(\mathbf{x}_n)$ results in a non-Gaussian posterior (not shown in Fig.\ \ref{fig:trajectories}).}
For the case of two targets as described in Section \ref{subsec:simu-settings}, Fig.\ \ref{fig:trajectories}(c) shows a realization of the 
target trajectories and the corresponding tracked trajectories that were obtained at one specific sensor by means of the LC-DPF. 
It can be seen that the target is tracked fairly well.
Other sensors obtained similar results. 

Table \ref{tab:compar} summarizes the estimation performance (ARMSE, track loss adjusted ARMSE, $\sigma_{\text{ARMSE}}$, track loss adjusted $\sigma_{\text{ARMSE}}$, and track loss percentage) and the communication requirements 
of the 
proposed consensus-based distributed PFs (LC-DPF, LC-DGPF, and R-LC-DGPF),
of the 
state-of-the-art consensus-based distributed PFs (GSHL-DPF, OC-DPF, and FRG-DPF),
and of the centralized methods (CPF and CGPF).
The ``track loss percentage'' is defined as the percentage of simulation runs during which the estimation error at time $n=200$ 
exceeded 5m, which is 
half the average inter-sensor distance. Such simulation runs were excluded in the calculation of the ``track loss adjusted'' 
$\text{RMSE}_n$,  
ARMSE, and $\sigma_{\text{ARMSE}}$. 
However, Table \ref{tab:compar} presents also the ARMSE and $\sigma_{\text{ARMSE}}$ 
computed using all the simulation runs (including those with lost tracks). 
The ``communication requirements'' are defined as the total number of real numbers transmitted (over one hop between neighboring sensors) at one time instant  
within the entire network. For the centralized methods (CPF and CGPF), we used multi-hop routing of measurements and sensor locations from every sensor to the FC (located in one of the corners of the network). Furthermore, the estimates calculated at the FC are disseminated throughout the network, such that every sensor obtains the centralized estimate. 

It is seen from Table \ref{tab:compar} that the track loss adjusted ARMSEs of the proposed distributed PFs are quite similar and that they are close to those of the centralized methods; 
they are slightly higher than that of FRG-DPF, slightly lower than that of OC-DPF, and 
about half that of GSHL-DPF. 
{For FRG-DPF, $\sigma_{\text{ARMSE}}$ is zero, 
since max and min consensus algorithms are employed to ensure 
identical 
results 
at each sensor. Furthermore, 
$\sigma_{\text{ARMSE}}$ is higher for LC-DPF and LC-DGPF than 
for R-LC-DGPF, GSHL-DPF, and OC-DPF. This is because 
R-LC-DGPF, GSHL-DPF, and OC-DPF 
employ a consensus step whereby 
Gaussian approximations of the partial/local posterior pdfs are combined to obtain a global posterior, thus achieving a tighter coupling between the sensors. 
By contrast, the local PFs of LC-DPF and LC-DGPF operate completely independently; 
only the JLF is computed in a distributed way using the LC scheme. 
Note, however, that the ARMSE and track loss adjusted ARMSE of LC-DPF and LC-DGPF are lower than for GSHL-DPF and OC-DPF.}  
Finally, the track loss percentages of the proposed distributed PFs are below 1\% and similar to those of GSHL-DPF, OC-DPF, and FRG-DPF. 
As a consequence, the ARMSEs are generally very close to the track loss adjusted ARMSEs.

The communication requirements of the distributed PFs are seen to be much higher than those of the centralized methods.
This is due to our low-dimensional (scalar) measurements and the fact that each local likelihood function is parametrized only by the sensor location, i.e., three real numbers must be transmitted in one hop. 
For high-dimensional measurements and/or a different parametrization of the local likelihood functions, resulting in about 190 or more real numbers to be transmitted in one hop, the opposite will be true. Note that 
even when 
the consensus-based methods require more communications, they may be preferable over centralized methods because
they are more robust (no possibility of FC failure), they require no routing protocols, and each sensor obtains an approximation of the global posterior 
(in the centralized schemes, each sensor obtains from the FC only the state estimate).
It is furthermore seen 
that the communication requirements of the proposed distributed PFs
are higher than those of GSHL-DPF and OC-DPF but much lower than those of FRG-DPF. 
Note, however, that the communication requirements of FRG-DPF depend on the number of particles and thus 
could be reduced by using fewer particles, whereas those of the other methods do not
depend on the number of particles. 
(A setting with a lower number of particles will be considered later.) 
Finally, among the proposed distributed PFs, the communication requirements of R-LC-DGPF
are higher by about 65\% than those of  LC-DPF and LC-DGPF.

\begin{figure*}[t]
\renewcommand{\baselinestretch}{1.7}\small\normalsize
\centering

\psfrag{n}[][][0.80]{\hspace{0mm}\raisebox{-10mm}{$n$\rule[-2.2mm]{0cm}{1cm}}}
\psfrag{RMSE}[][][0.80]{\hspace{0mm}\raisebox{8mm}{$\text{RMSE}_n$  [m]}}
\psfrag{TRMSE}[][][0.80]{\hspace{0mm}\raisebox{8mm}{Track loss adjusted $\text{RMSE}_n$  [m]}}

\psfrag{0.4}[][][0.6]{\hspace{-3mm}\raisebox{0mm}{$0.4$}}
\psfrag{0.6}[][][0.6]{\hspace{-3mm}\raisebox{0mm}{$0.6$}}
\psfrag{0.8}[][][0.6]{\hspace{-3mm}\raisebox{0mm}{$0.8$}}
\psfrag{1}[][][0.6]{\hspace{-3mm}\raisebox{0mm}{$1$}}
\psfrag{1.2}[][][0.6]{\hspace{-3mm}\raisebox{0mm}{$1.2$}}
\psfrag{1.4}[][][0.6]{\hspace{-3mm}\raisebox{0mm}{$1.4$}}
\psfrag{1.6}[][][0.6]{\hspace{-3mm}\raisebox{0mm}{$1.6$}}

\psfrag{0}[][][0.6]{\hspace{0mm}\raisebox{-4mm}{$0$}}
\psfrag{50}[][][0.6]{\hspace{0mm}\raisebox{-4mm}{$50$}}
\psfrag{100}[][][0.6]{\hspace{0mm}\raisebox{-4mm}{$100$}}
\psfrag{150}[][][0.6]{\hspace{0mm}\raisebox{-4mm}{$150$}}
\psfrag{200}[][][0.6]{\hspace{0mm}\raisebox{-4mm}{$200$}}

\hspace*{1.5mm}
\subfigure[ ]{

\psfrag{Gu-8iterCons}[][][0.80]{\hspace{7.8mm}\raisebox{0mm}{GSHL-DPF}}
\psfrag{Oreshkin-Coates-8iterCons}[][][0.80]{\hspace{-10mm}\raisebox{0mm}{OC-DPF}}
\psfrag{Farahmand-8iterCons}[][][0.80]{\hspace{-2.1mm}\raisebox{0mm}{FRG-DPF}}
\psfrag{DGPF-8iterCons}[][][0.80]{\hspace{2.7mm}\raisebox{0mm}{LC-DGPF}}
\psfrag{CGPF}[][][0.80]{\hspace{6.7mm}\raisebox{0mm}{CGPF}}

\includegraphics[height=5.1cm,width=7.1cm]{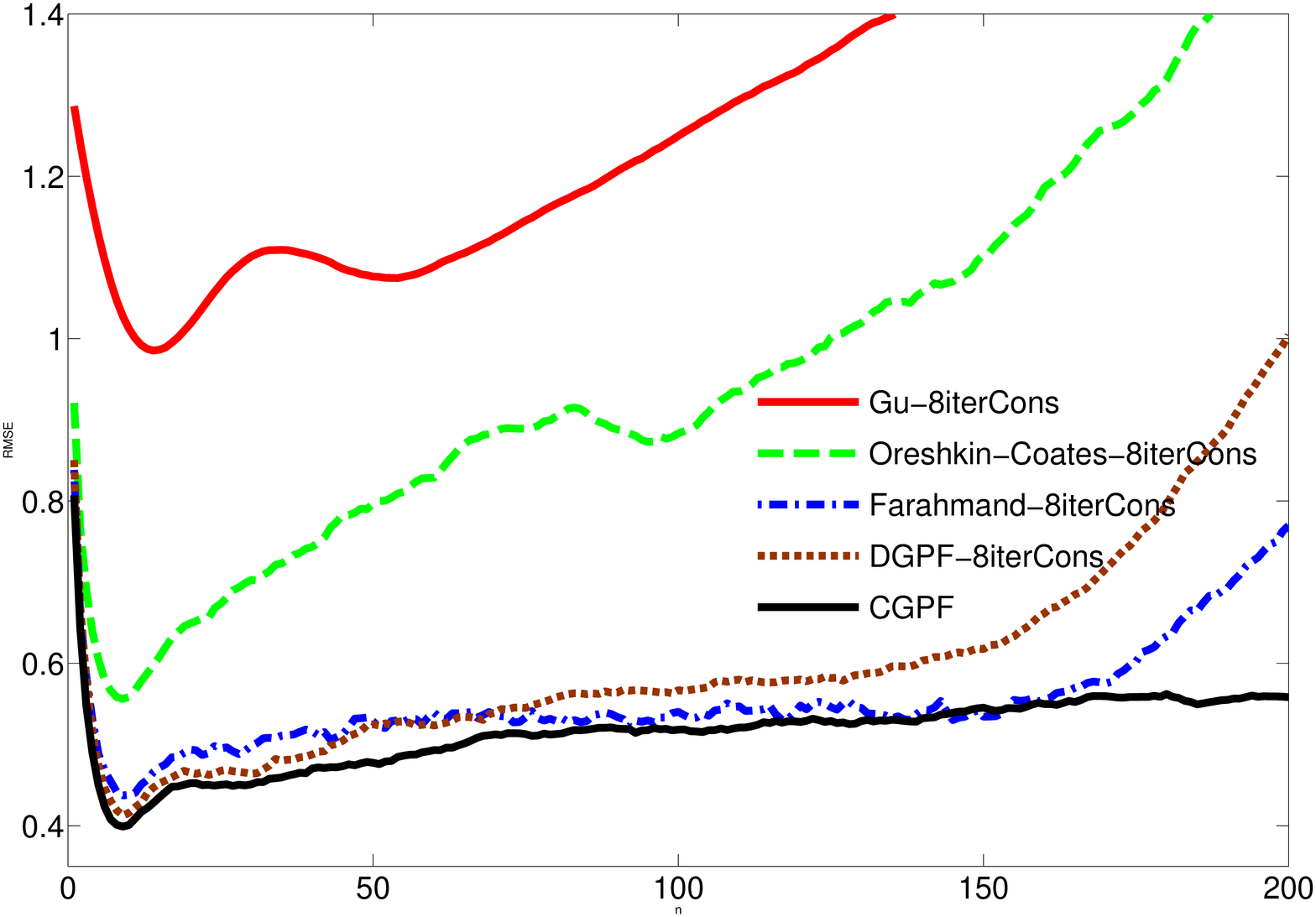}
}
\hspace{8mm}
\subfigure[ ]{

\psfrag{Gu-8iterCons}[][][0.80]{\hspace{7.8mm}\raisebox{0mm}{GSHL-DPF}}
\psfrag{Oreshkin-Coates-8iterCons}[][][0.80]{\hspace{-10mm}\raisebox{0mm}{OC-DPF}}
\psfrag{Farahmand-8iterCons}[][][0.80]{\hspace{-2.1mm}\raisebox{0mm}{FRG-DPF}}
\psfrag{DGPF-8iterCons}[][][0.80]{\hspace{2.7mm}\raisebox{0mm}{LC-DGPF}}
\psfrag{CGPF}[][][0.80]{\hspace{6.7mm}\raisebox{0mm}{CGPF}}

\includegraphics[height=5.1cm,width=7.1cm]{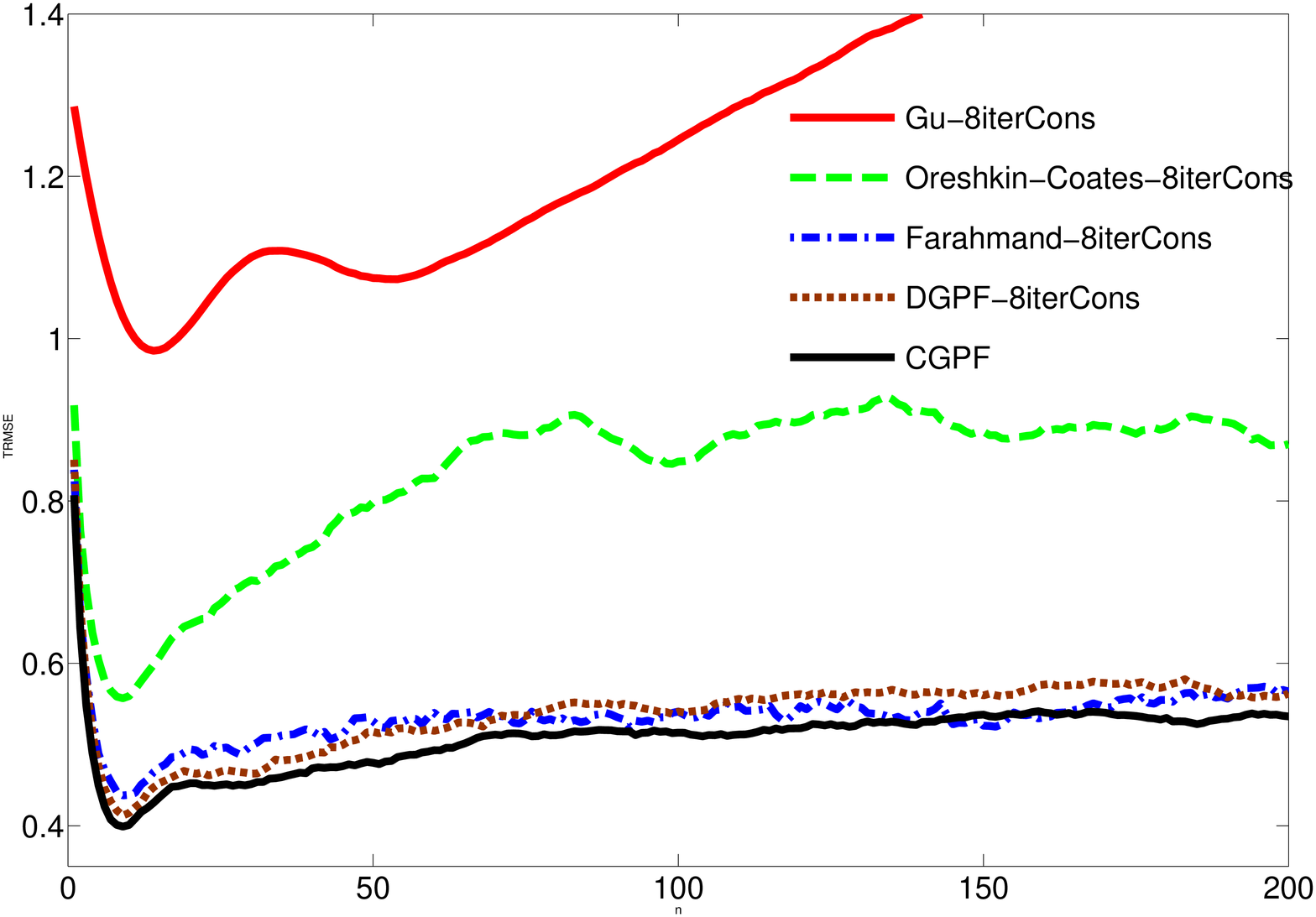}
}
\vspace*{-.5mm}
\renewcommand{\baselinestretch}{1.2}\small\normalsize
\caption{\label{fig:RMSEvsTime_otherMethods}{(a) RMSE$_n$} and (b) track loss adjusted RMSE$_n$
versus time $n$ for the proposed LC-DGPF, for the CGPF, and for state-of-the-art 
distributed PFs (GSHL-DPF, OC-DPF, and FRG-DPF). All distributed PFs use eight consensus iterations.}
\vspace{4.5mm}
\end{figure*}

In Fig.\ \ref{fig:RMSEvsTime_otherMethods}, we compare the  {$\text{RMSE}_n$ and track loss adjusted $\text{RMSE}_n$} of the proposed LC-DGPF with that of CGPF  
and the state-of-the-art 
distributed PFs (GSHL-DPF, OC-DPF, FRG-DPF). 
In terms of track loss adjusted $\text{RMSE}_n$ (Fig.\ \ref{fig:RMSEvsTime_otherMethods}(b)), 
LC-DGPF outperforms GSHL-DPF and OC-DPF, and it performs almost as well as FRG-DPF and CGPF. 
The increase in $\text{RMSE}_n$ over time in Fig.\ \ref{fig:RMSEvsTime_otherMethods}(a) is caused by the lost tracks. 

\begin{figure*}[t]
\renewcommand{\baselinestretch}{1.8}\small\normalsize
\centering

\psfrag{n}[][][0.80]{\hspace{0mm}\raisebox{-11mm}{$n$\rule[-2.2mm]{0cm}{1cm}}}
\psfrag{RMSE}[][][0.80]{\hspace{0mm}\raisebox{12mm}{$\text{RMSE}_n$  [m]}}
\psfrag{TRMSE}[][][0.80]{\hspace{0mm}\raisebox{12mm}{Track loss adjusted $\text{RMSE}_n$  [m]}}

\psfrag{0.4}[][][0.6]{\hspace{-5mm}\raisebox{0mm}{$0.4$}}
\psfrag{0.45}[][][0.6]{\hspace{-5mm}\raisebox{0mm}{$$}}
\psfrag{0.5}[][][0.6]{\hspace{-5mm}\raisebox{0mm}{$0.5$}}
\psfrag{0.55}[][][0.6]{\hspace{-5mm}\raisebox{0mm}{$$}}
\psfrag{0.6}[][][0.6]{\hspace{-5mm}\raisebox{0mm}{$0.6$}}
\psfrag{0.65}[][][0.6]{\hspace{-5mm}\raisebox{0mm}{$$}}
\psfrag{0.7}[][][0.6]{\hspace{-5mm}\raisebox{0mm}{$0.7$}}
\psfrag{0.75}[][][0.6]{\hspace{-5mm}\raisebox{0mm}{$$}}
\psfrag{0.8}[][][0.6]{\hspace{-5mm}\raisebox{0mm}{$0.8$}}
\psfrag{0.85}[][][0.6]{\hspace{-5mm}\raisebox{0mm}{$$}}

\psfrag{0}[][][0.6]{\hspace{0mm}\raisebox{-5mm}{$0$}}
\psfrag{50}[][][0.6]{\hspace{0mm}\raisebox{-5mm}{$50$}}
\psfrag{100}[][][0.6]{\hspace{0mm}\raisebox{-5mm}{$100$}}
\psfrag{150}[][][0.6]{\hspace{0mm}\raisebox{-5mm}{$150$}}
\psfrag{200}[][][0.6]{\hspace{0mm}\raisebox{-5mm}{$200$}}

\hspace*{1mm}
\subfigure[ ]{
\psfrag{LC-DPF-8iterCons}[][][0.80]{\hspace{35mm}\raisebox{0mm}{LC-DPF (8 consensus iterations)}}
\psfrag{LC-DPF-exactSum}[][][0.80]{\hspace{34mm}\raisebox{0mm}{LC-DPF (exact sum calculation)}}
\psfrag{CPF}[][][0.80]{\hspace{5mm}\raisebox{0mm}{CPF}}
\includegraphics[height=5cm,width=7cm]{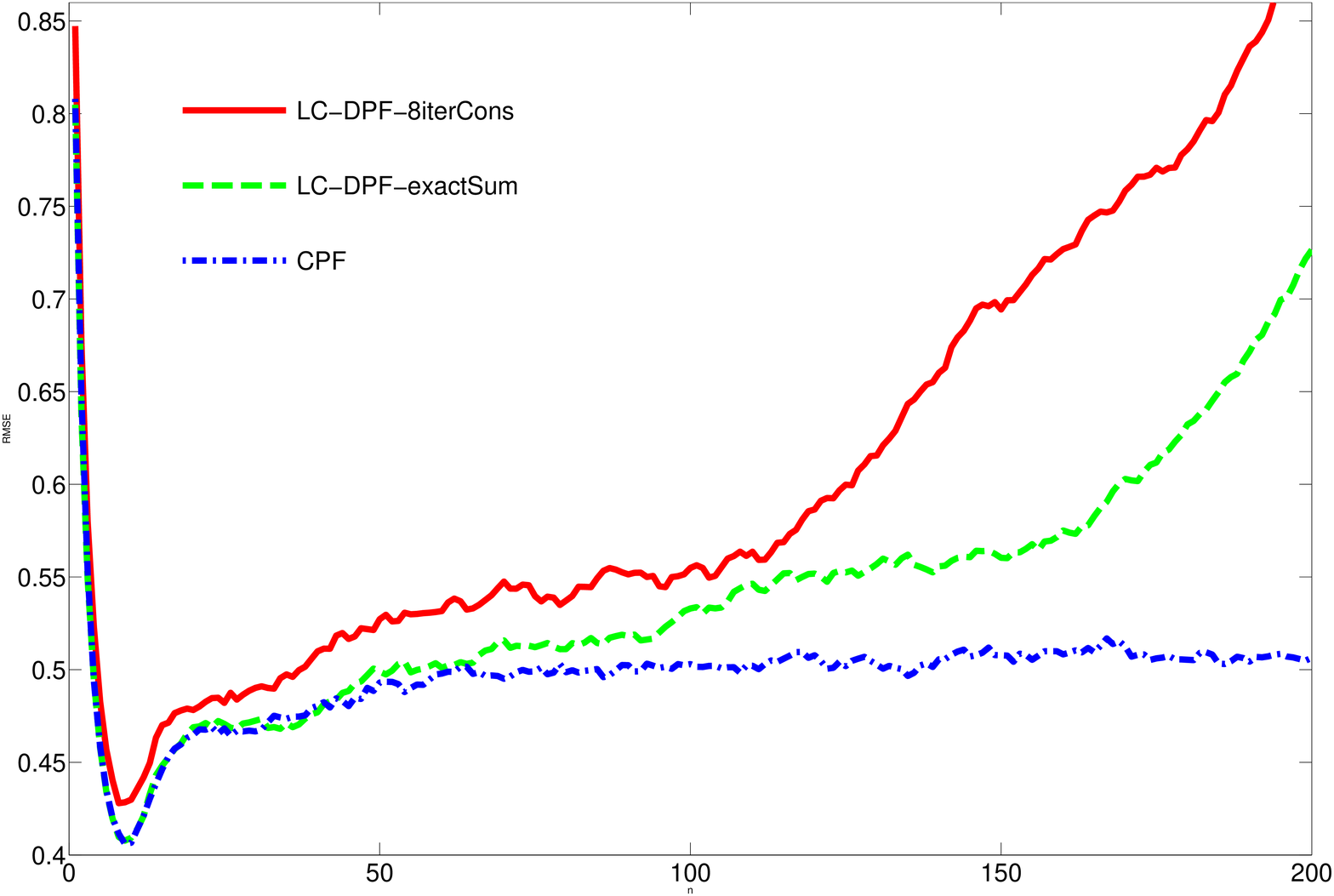}
}
\hspace{10mm}
\subfigure[ ]{
\psfrag{LC-DPF-8iterCons}[][][0.80]{\hspace{41.22mm}\raisebox{0mm}{LC-DPF (8 consensus iterations)}}
\psfrag{LC-DPF-exactSum}[][][0.80]{\hspace{40.5mm}\raisebox{0mm}{LC-DPF (exact sum calculation)}}
\psfrag{CPF}[][][0.80]{\hspace{6mm}\raisebox{0mm}{CPF}}
\includegraphics[height=5cm,width=7cm]{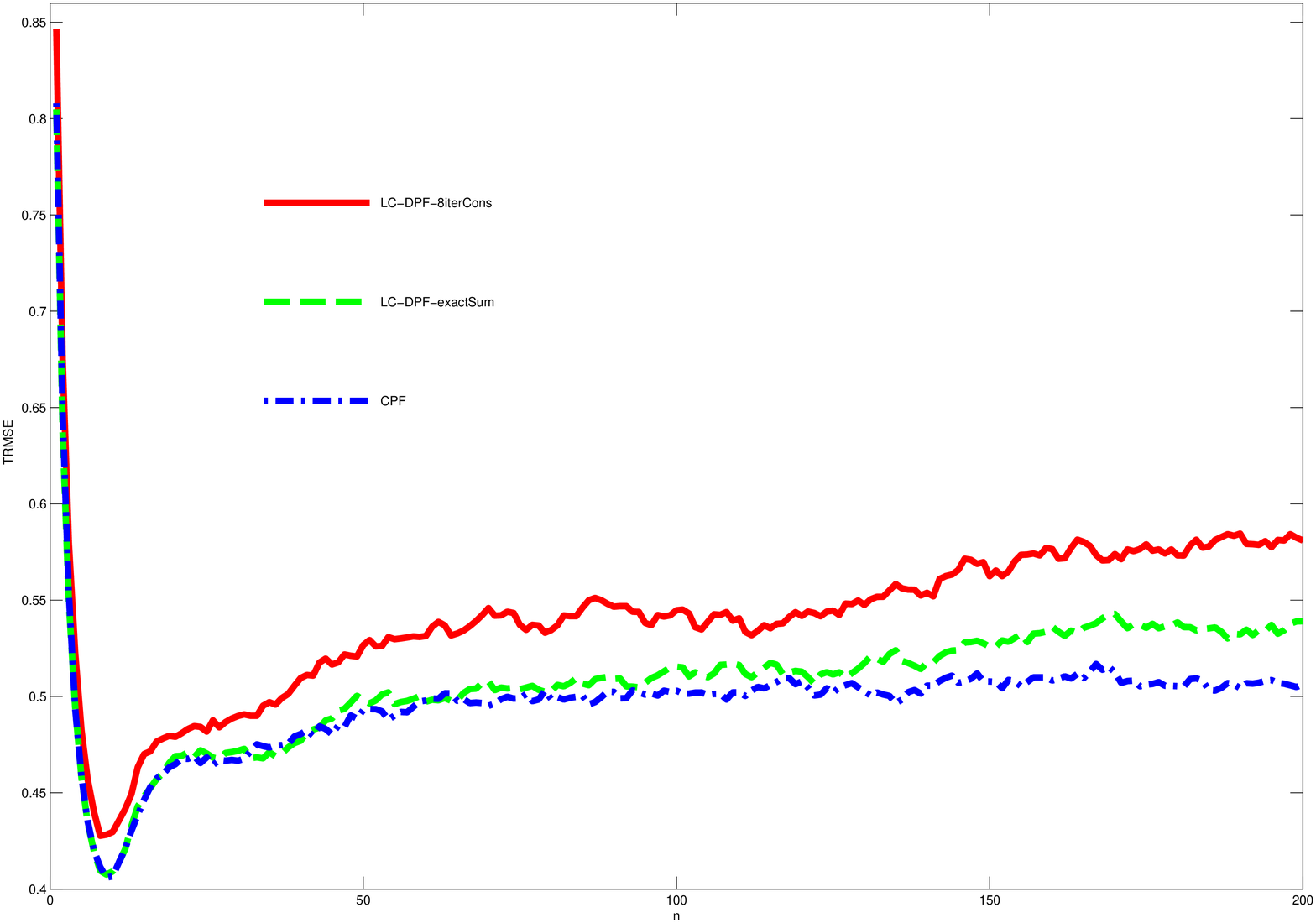}
}
\vspace*{-.5mm}
\renewcommand{\baselinestretch}{1.2}\small\normalsize
\caption{{(a) RMSE$_n$} and (b) track loss adjusted RMSE$_n$ 
versus time $n$ for the CPF, for the proposed LC-DPF using eight consensus iterations,
and for an impractical LC-DPF variant with exact sum calculation.}
\label{fig:RMSEvsTime_ConsVsExactSum}
\vspace*{-.5mm}
\end{figure*}

In Fig.\  \ref{fig:RMSEvsTime_ConsVsExactSum}, we compare the {RMSE$_n$ and track loss adjusted RMSE$_n$} 
of LC-DPF (using eight consensus iterations) with that of CPF. 
As a performance benchmark, we also show the results obtained by an impractical variant of LC-DPF in which the consensus algorithm is replaced by an 
exact, direct calculation of the sums in \eqref{eq:statistic_poly}. 
The performance degradation of LC-DPF with exact sum calculation relative to CPF is caused by the LS approximation of the
sensor measurement functions. The additional performance degradation of LC-DPF with eight consensus iterations relative to LC-DPF with exact sum calculation 
is due to the insufficiently converged consensus algorithms; it can be reduced by using more 
consensus iterations. In terms of the track loss adjusted RMSE$_n$, both performance degradations  are seen to be quite moderate. 
The track loss percentages were 0.95\% for LC-DPF, 0.29\% for LC-DPF with exact sum calculation, and 0\% for CPF.

\begin{figure}[t]
\vspace*{2.8mm}
\centering

\psfrag{Number of consensus iterations}[][][0.80]{\hspace{0mm}\raisebox{-6mm}{$I$}}
\psfrag{ARMSE [m]}[][][0.80]{\hspace{0mm}\raisebox{4mm}{Track loss adjusted ARMSE [m]}}

\psfrag{DGPF-exactSum}[][][0.80]{\hspace{21.0mm}\raisebox{0mm}{LC-DGPF (exact sum calculation)}}
\psfrag{DGPF-R-exactSum}[][][0.80]{\hspace{19.7mm}\raisebox{0mm}{R-LC-DGPF (exact sum calculation)}}
\psfrag{DGPF-withCons}[][][0.80]{\hspace{-12mm}\raisebox{0mm}{LC-DGPF}}
\psfrag{DGPF-R-withConsensus}[][][0.80]{\hspace{-23.5mm}\raisebox{0mm}{R-LC-DGPF}}

\psfrag{0.5}[][][0.6]{\hspace{-2mm}\raisebox{0mm}{$0.5$}}
\psfrag{0.55}[][][0.6]{\hspace{-2mm}\raisebox{0mm}{$0.55$}}
\psfrag{0.6}[][][0.6]{\hspace{-2mm}\raisebox{0mm}{$0.6$}}
\psfrag{0.65}[][][0.6]{\hspace{-2mm}\raisebox{0mm}{$0.65$}}
\psfrag{0.7}[][][0.6]{\hspace{-2mm}\raisebox{0mm}{$0.7$}}
\psfrag{0.75}[][][0.6]{\hspace{-2mm}\raisebox{0mm}{$0.75$}}

\psfrag{4}[][][0.6]{\hspace{0mm}\raisebox{0mm}{$4$}}
\psfrag{6}[][][0.6]{\hspace{0mm}\raisebox{0mm}{$6$}}
\psfrag{8}[][][0.6]{\hspace{0mm}\raisebox{0mm}{$8$}}
\psfrag{10}[][][0.6]{\hspace{0mm}\raisebox{0mm}{$10$}}
\psfrag{12}[][][0.6]{\hspace{0mm}\raisebox{0mm}{$12$}}
\psfrag{14}[][][0.6]{\hspace{0mm}\raisebox{0mm}{$14$}}
\psfrag{16}[][][0.6]{\hspace{0mm}\raisebox{0mm}{$16$}}

\hspace*{-3mm}\includegraphics[height=5.4cm,width=7.6cm]{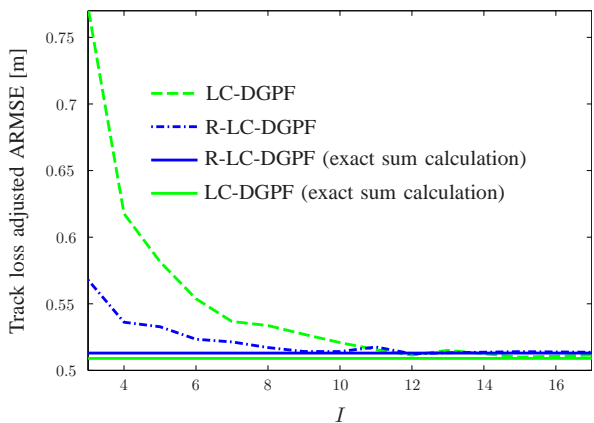}
\vspace*{0mm}
\renewcommand{\baselinestretch}{1.2}\small\normalsize
\caption{Track loss adjusted ARMSE of the LC-DGPF and R-LC-DGPF versus the number $I$ of consensus iterations, along with the 
track loss adjusted ARMSE of the impractical LC-DGPF and R-LC-DGPF variants with exact sum calculation.
(R-LC-DGPF uses $I$ consensus iterations for each sum calculation.)}
\label{fig:ARMSEvsIter}
\vspace{-1mm}
\end{figure}

Fig.\ \ref{fig:ARMSEvsIter} shows 
the track loss adjusted ARMSE of the proposed LC-DGPF and R-LC-DGPF versus 
the number $I$ of consensus iterations. Here, R-LC-DGPF uses $I$ consensus iterations in each 
one of its two consensus stages 
(i.e., $I$ iterations to compute the sums in \eqref{eq:statistic_poly} and $I$ iterations 
each 
to compute the sums in \eqref{eq:muAndC_complete} and \eqref{eq:W_complete}).
As a performance benchmark, the figure also shows the results for impractical variants of LC-DGPF and R-LC-DGPF using exact, direct calculation of the sums 
\eqref{eq:statistic_poly}, \eqref{eq:muAndC_complete}, and \eqref{eq:W_complete}.
It is seen that the performance of the impractical direct calculation is essentially achieved for $I$ about 7 in the case of R-LC-DGPF 
and for $I$ about 10 in the case of LC-DGPF. Somewhat surprisingly, R-LC-DGPF outperforms LC-DGPF for up to 10 consensus 
iterations, i.e., the additional consensus algorithms 
used to calculate the sums in \eqref{eq:muAndC_complete} and \eqref{eq:W_complete} result in a better performance of R-LC-DGPF,
in spite of the significantly reduced number of particles (200 instead of 5000). 
However, as the number of consensus iterations increases, 
both methods approach the performance of the respective ``exact sum calculation'' variant and LC-DGPF slightly outperforms R-LC-DGPF. 
This behavior can be explained as follows. 
The LC with a small number of consensus iterations is not completely converged, which means that the local information is not yet completely diffused 
throughout the network and the resulting approximate JLF does not yet contain the complete global information. 
The additional consensus stage of R-LC-DGPF then helps
to further diffuse the local information. 

Finally, we consider a setting where each sensor in the distributed PF methods (LC-DPF, LC-DGPF, GSHL-DPF, OC-DPF, and FRG-DPF) 
as well as the FC in 
CPF and CGPF use only $J=400$ particles, 
and consequently R-LC-DGPF uses only $J'=400/25=16$ particles per sensor. 
This reduction of the number of particles results in reduced communication requirements of FRG-DPF but not of the other methods as their 
communication requirements are independent of the number of particles. Table \ref{tab:compar2} summarizes the simulation results we obtained. 
A comparison with Table \ref{tab:compar} shows that, as expected, the performance of all methods is degraded. 
Furthermore, the high ARMSE and track loss percentage values of LC-DPF, LC-DGPF, and OC-DPF can be viewed as signs of divergence. 
In the case of LC-DPF and LC-DGPF, high $\sigma_{\text{ARMSE}}$ values indicate significant differences between the local particle representations of the global posterior; 
these differences reduce the effectiveness of the LS approximation in the LC scheme. 
In the case of OC-DPF, the divergence is due to the peaky functions (powers of local likelihoods functions) used in the weight update, which
cause most of the particles to be located 
in regions of low 
likelihood. 
FRG-DPF performs well due to its use of adapted proposal distributions; its communication requirements are now closer to those of 
the other methods but still 
higher. 
R-LC-DGPF is seen to perform even slightly better with, at the same time, lower communication costs. As mentioned before, the additional consensus algorithms used by
R-LC-DGPF  
lead to very similar particle representations of the local PFs across the network, with particles located in almost identical regions of the state space;
this is evidenced by the low value of $\sigma_{\text{ARMSE}}$. 
Therefore, all sensors perform the LS approximation of their local likelihood functions in almost the same state space region,
which moreover is the region where the particles of \emph{all} sensors are located.
Combining the local approximations using the LC scheme, we thus obtain a JLF approximation that is most accurate in that state space region. 
This explains the good tracking performance of R-LC-DGPF.

\begin{table*}[t]
\centering
{
\begin{tabular}{|c||c|c|c|c|c|c|}
  \hline
&  & \rule{0mm}{3.2mm}Track loss adjusted &  & Track loss adjusted & Track loss & Communication    \\[-.01mm]
& ARMSE [m] & ARMSE [m] & $\sigma_{\text{ARMSE}}\,$[m] & $\sigma_{\text{ARMSE}}\,$[m]  & percentage [\%] &  requirements   \\[.8mm]
  \hline \hline
  \rule{0mm}{3mm}LC-DPF & 3.3137 & 0.9651 & 0.1882 & 0.1523 & 26.34 & 13800 \\[.2mm]
  \hline
  \rule{0mm}{3mm}LC-DGPF & 1.8692 & 0.6704 & 0.1074 & 0.0512 & 5.68 & 13800 \\[.2mm]
  \hline
  \rule{0mm}{3mm}R-LC-DGPF & 0.7600 & 0.6556 & 0.0005 & 0.0004 & 0.71 & 22800 \\[.2mm]
  \hline
  \rule{0mm}{3mm}GSHL-DPF \cite{gu2008consensus} & 1.3371 & 1.3200 & 0.0032 & 0.0032 & 0.42 & 8800 \\[.2mm]
  \hline
  \rule{0mm}{3mm}OC-DPF \cite{oreshkin2010async} & 2.9276 & 1.7746 & 0.0008 & 0.0015 & 28.50 & 8800 \\[.2mm]
  \hline
  \rule{0mm}{3mm}FRG-DPF \cite{farahmand2011set} & 0.8587 & 0.7162 & 0 & 0 & 1.01 & 80000 \\[.2mm]
  \hline     
  \rule{0mm}{3mm}CPF & 1.0552 & 0.7300 & -- & -- & 2.48 & 770 \\[.2mm]
  \hline
  \rule{0mm}{3mm}CGPF & 0.6702 & 0.5879 & -- & -- & 0.44 & 770 \\[.2mm]
  \hline
\end{tabular}
}
\vspace{2mm}
\renewcommand{\baselinestretch}{.9}\small\normalsize
\caption{{Same comparison as in Table \ref{tab:compar}, but using only $J=400$ particles per sensor for LC-DPF, LC-DGPF, GSHL-DPF, OC-DPF, and FRG-DPF;
$J'=16$ particles per sensor for R-LC-DGPF;
and $400$ particles at the fusion center for CPF and CGPF.}} 
\label{tab:compar2}
\vspace*{5mm}
\end{table*}

\section{Conclusion}\label{sec:conclusion}

For global estimation tasks in wireless sensor networks, the joint (all-sensors) likelihood function (JLF) plays a central role because it epitomizes the measurements of all sensors. We proposed a distributed, consensus-based method for computing the JLF. This ``likelihood consensus'' method uses iterative consensus algorithms to compute, at each sensor, an approximation of the JLF as a function of the state to be estimated. 
Our method is applicable if the local likelihood functions of the various sensors (viewed as 
conditional probability density functions of the local measurements)
belong to the exponential family of distributions. This includes the case of additive Gaussian measurement noises. 
The employed consensus algorithms require only 
local communications between neighboring sensors and operate without complex routing protocols. 

We demonstrated the use of the likelihood consensus method for distributed particle filtering and distributed Gaussian particle filtering.
At each sensor, a local particle filter computes a global state estimate that reflects the measurements of all sensors. The approximate JLF provided by 
the likelihood consensus method is used for updating
the particle weights of each local particle filter. A second stage of consensus algorithms can be employed to significantly reduce 
the complexity of the distributed Gaussian particle filter. 
We applied the proposed distributed particle filters to a multiple target 
tracking problem and demonstrated experimentally that their performance is close to that of the centralized particle filters. 
Compared to three state-of-the-art distributed particle filtering schemes, our methods typically achieve a comparable or 
better estimation performance, while the communication requirements are somewhat higher in two cases and much lower in one case. 

We finally note that the proposed distributed Gaussian particle filter can be extended to a consensus-based, distributed
implementation of the Gaussian sum particle filter proposed in \cite{kotecha2003gaussiansum}. Furthermore, an extension of the likelihood consensus 
method to general local likelihood functions (i.e., not necessarily belonging to the exponential family)
has been presented in \cite{hlinka2012prop}.

\vspace*{-2mm}

\bibliographystyle{ieeetr}


\begin{biography}[{\includegraphics[width=1in,height=1.25in,clip,keepaspectratio]{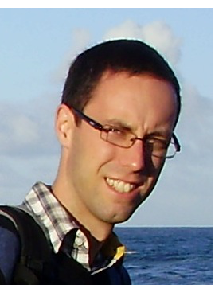}}]{Ondrej Hlinka} 
(S'09) received the Ing.\ (M.~Eng.)\ degree in electrical engineering from the Slovak University of Technology, Bratislava, Slovakia. 
Since 2008, he has been a Research Assistant with the Institute of Telecommunications, Vienna University of Technology, Vienna, Austria,
where he is working toward the Ph.D. degree. His research interests include signal processing for wireless sensor networks, statistical signal processing, and sequential Monte Carlo methods.  
\end{biography}

\begin{biography}[{\includegraphics[width=1in,height=1.25in,clip,keepaspectratio]{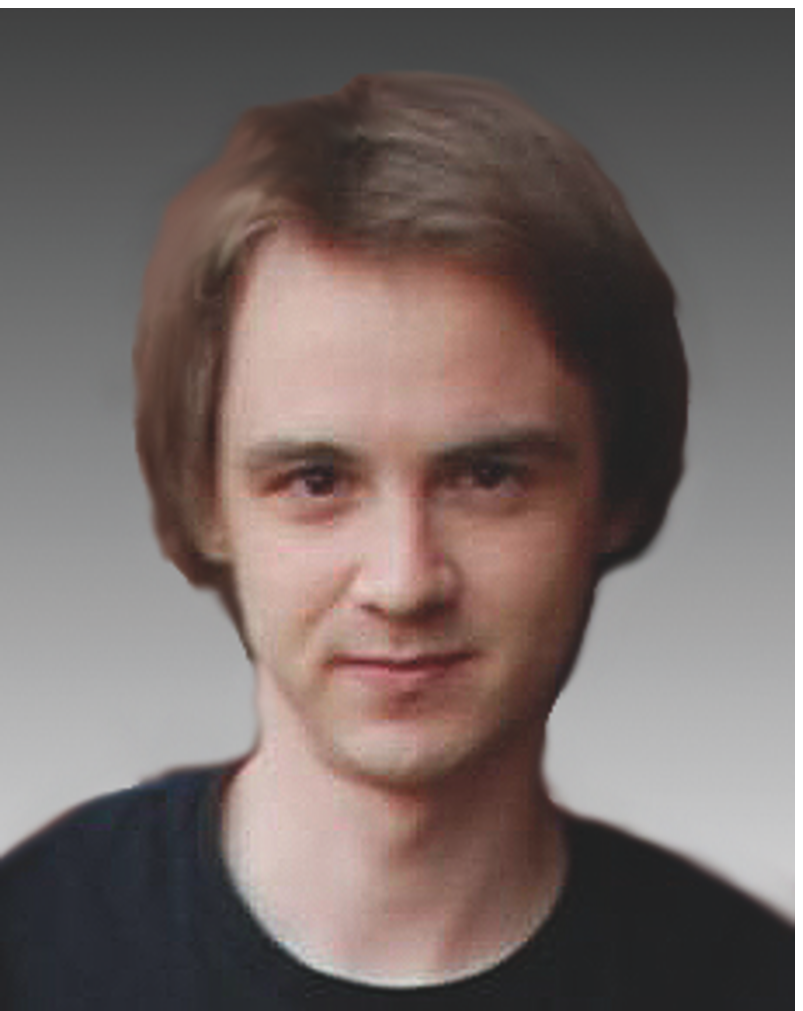}}]{Ondrej Slu\v ciak} 
(S'10) received the Ing.\ (M.~Eng.)\ degree in electrical engineering from the Slovak University of Technology, Bratislava, Slovakia.
Since 2008, he has been a Research Assistant with the Institute of Telecommunications, Vienna University of Technology, Vienna, Austria,
where he is working toward the Ph.D. degree. His research interests include signal processing for wireless sensor networks, distributed consensus algorithms, and their convergence analysis.  
\end{biography}

\begin{biography}[{\includegraphics[width=1in,height=1.25in,clip,keepaspectratio]{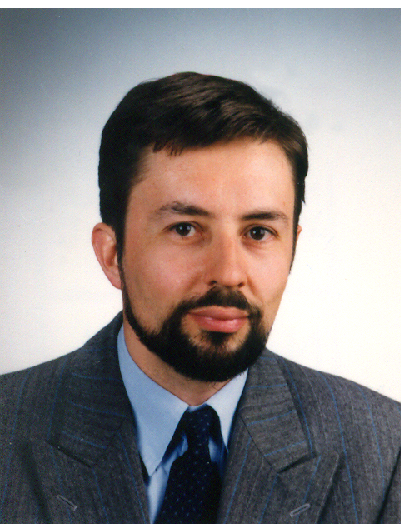}}]{Franz Hlawatsch} 
(S'85--M'88--SM'00--F'12) received the Diplom-Ingenieur, Dr. techn., and Univ.-Dozent (habilitation) degrees in electrical engineering/signal processing from Vienna University of Technology, Vienna, Austria in 1983, 1988, and 1996, respectively.

Since 1983, he has been with the Institute of Telecommunications, Vienna University of Technology, where he is currently an Associate Professor. During 1991--1992, as a recipient of an Erwin Schr\"odinger Fellowship, he spent a sabbatical year with the Department of Electrical Engineering, University of Rhode Island, Kingston, RI, USA. In 1999, 2000, and 2001, he held one-month Visiting Professor positions with INP/ENSEEIHT, Toulouse, France and IRCCyN, Nantes, France. He (co)authored a book, a review paper that appeared in the {\sc IEEE Signal Processing Magazine}, about 190 refereed scientific papers and book chapters, and three patents. He coedited three books. His research interests include signal processing for wireless communications and sensor networks, statistical signal processing, and compressive signal processing.

Prof. Hlawatsch was Technical Program Co-Chair of EUSIPCO 2004 and served on the technical committees of numerous IEEE conferences. He was an Associate Editor for the {\sc IEEE Transactions on Signal Processing} from 2003 to 2007 and for the {\sc IEEE Transactions on Information Theory} from 2008 to 2011. From 2004 to 2009, he was a member of the IEEE SPCOM Technical Committee. He is coauthor of papers that won an IEEE Signal Processing Society Young Author Best Paper Award and a Best Student Paper Award at IEEE ICASSP 2011.
\end{biography}

\begin{biography}[{\includegraphics[width=1in,height=1.25in,clip,keepaspectratio]{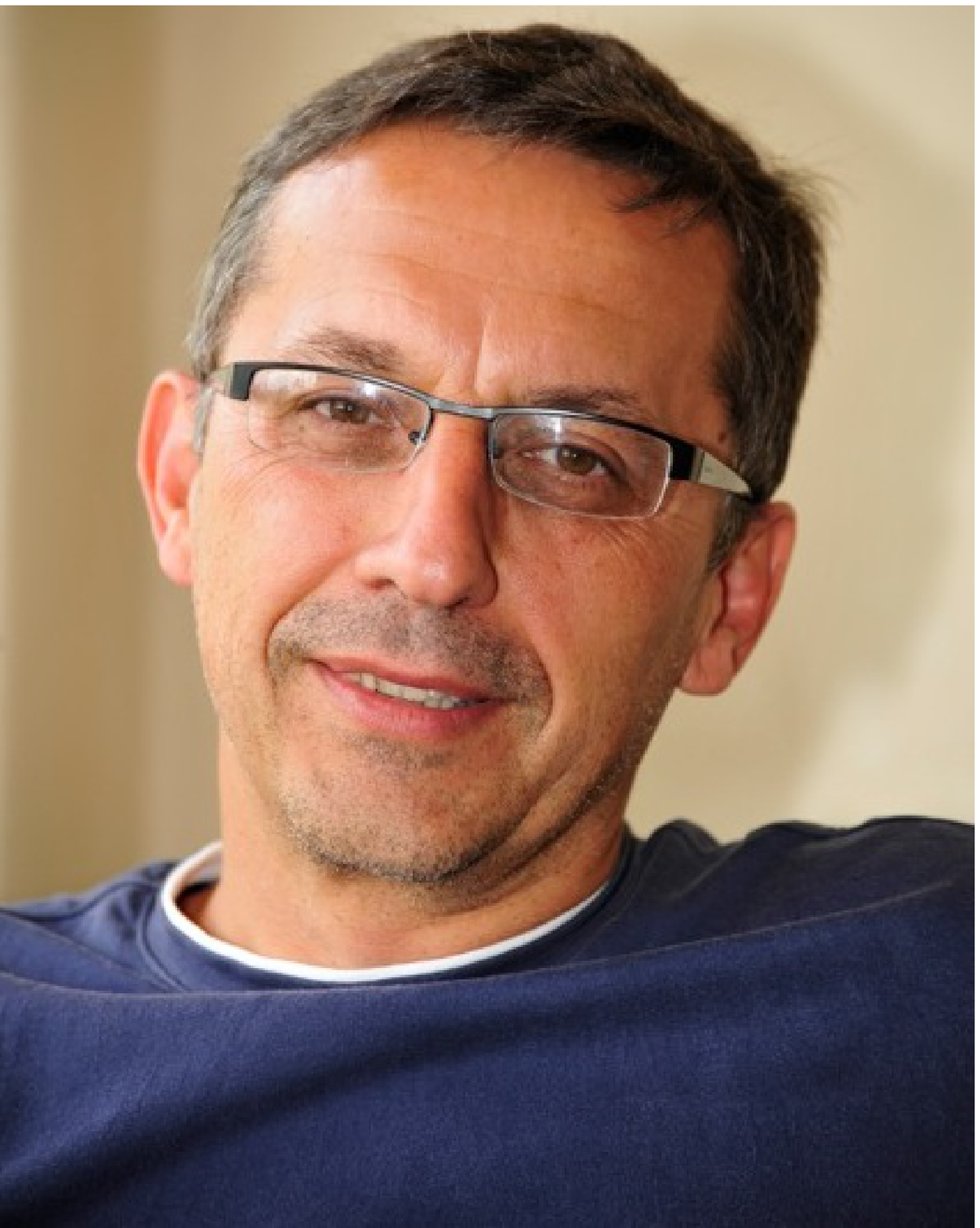}}]{Petar M.\ Djuri{\'c}} (S'86--M'90--SM'99--F'06)
received his B.S.\ and M.S.\ degrees in electrical engineering from the University of Belgrade in 1981 and 1986, respectively, and his Ph.D.\ degree in electrical engineering from the University of Rhode Island in 1990. 

From 1981 to 1986, he was a Research Associate with the Institute of Nuclear Sciences, Vin{\v c}a, Belgrade. Since 1990, he has been with Stony Brook University, where he is currently a Professor in the Department of Electrical and Computer Engineering. He works in the area of statistical signal processing, and his primary interests are in the theory of signal modeling, detection, and estimation and application of the theory to a wide range of disciplines. 

Prof. Djuri{\'c} has been invited to lecture at many universities in the United States and overseas. In 2007, he received the {\sc IEEE Signal Processing Magazine} Best Paper Award, and in 2008, he was elected Chair of Excellence of Universidad Carlos III de Madrid-Banco de Santander. During 2008--2009, he was Distinguished Lecturer of the IEEE Signal Processing Society. In 2012, he received the EURASIP Technical Achievement Award. He has served on numerous committees for the IEEE, and currently he is a Member-at-Large of the Board of Governors of the Signal Processing Society. 
\end{biography}

\begin{biography}[{\includegraphics[width=1in,height=1.25in,clip,keepaspectratio]{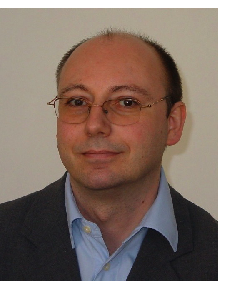}}]{Markus Rupp} (M'03--SM'06) 
received the Dipl.-Ing. degree at the University of Saarbr\"ucken, Germany, in 
1988 and the Dr.-Ing. degree at the Technische Universit\"at 
Darmstadt, Germany, in 1993, where he worked 
with E. H\"ansler on designing new algorithms 
for acoustical and electrical echo compensation. 
From November 1993 until July 1995, he had
a postdoctoral position at the University of Santa 
Barbara, Santa Barbara, CA, with S. Mitra, where he 
worked with A. H. Sayed on a robustness description 
of adaptive filters with impact on neural networks 
and active noise control. From October 1995 until August 2001, he was a 
member of Technical Staff in the Wireless Technology Research Department
of Bell Labs, Crawford Hill, NJ, where he worked on various topics related to 
adaptive equalization and rapid implementation for IS-136, 802.11 and UMTS,
including the first MIMO prototype for UMTS. Since October 2001, he has
been a Full Professor for Digital Signal Processing in Mobile Communications 
at the Vienna University of Technology, Vienna, Austria, where he founded the
Christian-Doppler Laboratory for Design Methodology of Signal Processing 
Algorithms in 2002 at the Institute for Communications and Radio-Frequency Engineering.
He served as Dean during 2005--2007. He authored and coauthored more than
400 scientific papers and patents on adaptive filtering, wireless communications,
and rapid prototyping, as well as automatic design methods.

Prof. Rupp was Associate Editor of the IEEE {\sc Transactions on Signal Processing} during 2002--2005, 
and is currently Associate Editor of the \emph{EURASIP Journal of Advances in Signal Processing} and the \emph{EURASIP Journal on
Embedded Systems}. He has been an elected AdCom member of EURASIP
since 2004 and served as President of EURASIP during 2009--2010.
\end{biography}

\onecolumn

\clearpage\newpage

\end{document}